\documentclass[longauth,referee]{aa} 

\usepackage{graphicx}
\usepackage{lscape}
\usepackage{txfonts}
\usepackage{subcaption}
\usepackage{lscape}
\usepackage{color, soul} 
\usepackage{sidecap}
\newcommand{\msun}{M_\odot}

\def\ms{\hbox{\,m\,s$^{-1}$}}         
\def\kms{\hbox{\,km\,s$^{-1}$}}       
\def\sini{\hbox{sin\,$i$}}      
\def\Rsun{\hbox{$\mathrm{R}_{\odot}$}}
\def\Mjup{\hbox{$\mathrm{M}_{\rm Jup}$}}
\def\Rjup{\hbox{$\mathrm{R}_{\rm Jup}$}}

\begin{document} 

   \title{The GAPS Programme at TNG}
   
   \subtitle{XXVII. Reassessment of a young planetary system with HARPS-N: is the hot Jupiter \object{\object{V830 Tau}}\,b really there?\thanks{Based on observations made with the Italian Telescopio Nazionale Galileo (TNG) operated on the island of La Palma by the Fundaci\'on Galileo Galilei of the INAF (Istituto Nazionale di Astrofisica) at the Spanish Observatorio del Roque de los Muchachos of the IAC. Tables A.1, A.2, B.1, and C.1 are only available in electronic form at the CDS via anonymous ftp to cdsarc.u-strasbg.fr (130.79.128.5) or via http://cdsweb.u-strasbg.fr/cgi-bin/qcat?J/A+A/}}

   \author{M.~Damasso, \inst{1}
            A.~F.~Lanza, \inst{2} 
            S.~Benatti, \inst{3}     
            V.~M.~Rajpaul, \inst{4} 
            M.~Mallonn, \inst{5} 
            S.~Desidera, \inst{6}
            K.~Biazzo, \inst{7}
            V.~D'Orazi, \inst{6}
            L.~Malavolta, \inst{8}
            D.~Nardiello, \inst{9,6}
            M.~Rainer, \inst{10}
            F.~Borsa, \inst{11}
            L.~Affer, \inst{3}
            A.~Bignamini, \inst{12} 
            A.S.~Bonomo,  \inst{1} 
            I.~Carleo, \inst{13}
            R.~Claudi, \inst{6}
            R.~Cosentino, \inst{14}
            E.~Covino, \inst{14}
            P.~Giacobbe,  \inst{1} 
            R.~Gratton, \inst{6}
            A.~Harutyunyan, \inst{14}
            C.~Knapic, \inst{12}
            G.~Leto, \inst{2}
            A.~Maggio, \inst{3}
            J.~Maldonado, \inst{3}
            L.~Mancini, \inst{16,17,1}
            G.~Micela, \inst{3}
            E.~Molinari, \inst{18}
            V.~Nascimbeni, \inst{6}
            I.~Pagano, \inst{2}
            G.~Piotto, \inst{8}
            E.~Poretti, \inst{14}
            G.~Scandariato, \inst{2}
            A.~Sozzetti,  \inst{1} 
            R.~Capuzzo Dolcetta, \inst{19}
            M.P.~Di Mauro, \inst{20}
            D.~Carosati, \inst{14}
            A.~Fiorenzano, \inst{14}
            G.~Frustagli, \inst{11,21}
            M.~Pedani, \inst{14}
            M.~Pinamonti, \inst{1}
            H.~Stoev, \inst{14}
            D.~Turrini, \inst{20}
           }

   \institute{INAF -- Osservatorio Astrofisico di Torino, Via Osservatorio 20, I-10025, Pino Torinese (TO), Italy 
              \email{mario.damasso@inaf.it}
               \and INAF -- Osservatorio Astrofisico di Catania, Via S.~Sofia,78 - 95123 Catania, Italy
               \and INAF -- Osservatorio Astronomico di Palermo, Piazza del Parlamento 1, I-90134, Palermo, Italy
               \and Astrophysics Group, Cavendish Laboratory, JJ Thomson Avenue, CB3 0HE Cambridge, UK
               \and Leibniz-Institut f\"ur Astrophysik Potsdam, An der Sternwarte 16, 14482 Potsdam, Germany
               \and INAF -- Osservatorio Astronomico di Padova, Vicolo dell'Osservatorio 5, IT-35122, Padova, Italy
               \and INAF -- Osservatorio Astronomico di Roma, Via Frascati 33, I-00040, Monte Porzio Catone (RM), Italy
               \and Dipartimento di Fisica e Astronomia "G. Galilei"-- Universt\`a degli Studi di Padova, Vicolo dell'Osservatorio 3, I-35122 Padova
               \and Aix Marseille Univ, CNRS, CNES, LAM, Marseille, France
               \and INAF -- Osservatorio Astrofisico di Arcetri, Largo Enrico Fermi, 5, I-50125 Firenze, Italy 
               \and INAF -- Osservatorio Astronomico di Brera, Via E. Bianchi 46, I-23807 Merate (LC), Italy
               \and INAF -- Osservatorio Astronomico di Trieste, via Tiepolo 11, I-34143 Trieste, Italy
               \and Astronomy Department and Van Vleck Observatory, Wesleyan University, Middletown, CT 06459, USA
               \and Fundación Galileo Galilei - INAF, Rambla Jos\'e Ana Fernandez P\'erez 7, E-38712, Bre\~na Baja, TF - Spain
               \and INAF -- Osservatorio Astronomico di Capodimonte, Salita Moiariello 16, I-80131 Napoli, Italy
               \and Department of Physics, University of Rome Tor Vergata, Via della Ricerca Scientifica 1 I-00133 Roma, Italy
               \and Max Planck Institute for Astronomy, K\"{o}nigstuhl 17, D-69117, Heidelberg, Germany
               \and INAF -- Osservatorio Astronomico di Cagliari, Via della Scienza 5, I-09047, Selargius (CA), Italy               
               \and Dipartimento di Fisica, Universit\`a di Roma La Sapienza, P.le A.Moro 5, I-00185, Roma, Italy
               \and INAF -- Istituto di Astrofisica e Planetologia Spaziali, Via del Fosso del Cavaliere 100, I-00133, Roma, Italy
               \and Dipartimento di Fisica G. Occhialini, Università degli Studi di Milano-Bicocca, Piazza della Scienza 3, I-20126 Milano, Italy
             }

   \date{Received XXX 2020; accepted YYY }

 
  \abstract
   {Detecting and characterising exoworlds around very young stars (age $\leqslant$10 Myr) are key aspects of exoplanet demographic studies, especially for understanding the mechanisms and timescales of planet formation and migration. Actually, any reliable theory for such physical phenomena requires a robust observational data base to be tested. However, detection using the radial velocity method alone can be very challenging, since the amplitude of the signals due to magnetic activity of such stars can be orders of magnitude larger than those induced even by massive planets.}
   {We observed the very young ($\sim$2 Myr) and very active star \object{V830 Tau} with the HARPS-N spectrograph between Oct 2017 and Mar 2020 to independently confirm and characterise the previously reported hot Jupiter \object{V830 Tau~b} ($K_{\rm b}=68\pm11$~$\ms$; $m_{\rm b}\sini_{\rm b}=0.57\pm0.10$ $\Mjup$; $P_{\rm b}=4.927\pm0.008$~d).}
   {Due to the observed $\sim$1 $\kms$ radial velocity scatter clearly attributable to V830 Tau's magnetic activity, we analysed radial velocities extracted with different pipelines and modelled them using several state-of-the-art tools. We devised injection-recovery simulations to support our results and characterise our detection limits. The analysis of the radial velocities was aided by a characterisation of the stellar activity using simultaneous photometric and spectroscopic diagnostics.}
   {Despite the high quality of our HARPS-N data and the diversity of tests we performed, we could not detect the planet \object{V830 Tau~b} in our data and confirm its existence. Our simulations show that a statistically-significant detection of the claimed planetary Doppler signal is very challenging.}
   {Much as it is important to continue Doppler searches for planets around young stars, utmost care must be taken in the attempt to overcome the technical difficulties to be faced in order to achieve their detection and characterisation. This point must be kept in mind when assessing their occurrence rate, formation mechanisms and migration pathways, especially without evidence of their existence from photometric transits.}

\titlerunning{The GAPS programme: no planet around \object{V830 Tau}?}
\authorrunning{M. Damasso et al.}
\maketitle
%
\section{Introduction}
\label{sec:intro}
Exoplanetary systems known to date show a large variety of architectures, resulting from the diverse outcomes of planet formation and evolution processes. 
Planetary migration mechanisms are acknowledged to be the main factor responsible for shaping the observed systems, and possibly for the origin of the hot Jupiters (HJs, e.g. \citealt{2018ARA&A..56..175D}). Planet-disk interaction \citep{2014prpl.conf..667B}, high-eccentricity migration \citep{1996Sci...274..954R} produced by secular interaction among bodies in the system or planet-planet scattering, or in-situ formation \citep{2016ApJ...829..114B} are expected to produce observable trends in the planet population that can be used to gauge their respective effectiveness. Theoretical works partially describe the observed distribution of the HJ population \citep{2008ApJ...686..621F,2010ApJ...725.1995M,2017MNRAS.464..688H}, which seems to be mainly produced by a high-eccentricity migration process associated with tidal interactions (e.g., \citealt{2017A&A...602A.107B}). However, this scenario cannot fully explain the observational evidence, and a clear view of the conditions that favour one mechanism over the others is still missing. Information on the HJ formation path can be obtained from determination of their orbital parameters, in particular eccentricity and obliquity, but also from the understanding of the migration timescales and age-dependent frequency of different types of systems. However, these clues cannot be easily provided by the available and well-known distribution of mature systems. 
Instead, observation of HJs around young stars allows to directly spot the ongoing planetary evolution and provide crucial indications to this open question.

In recent years, the first detections of exoplanets in young open clusters and stellar associations have been claimed (e.g., \citealt{quinn12,quinn14,2016A&A...588A.118M,2017AJ....153...64M,2018AJ....155....4M}).
One of the most intriguing results is the apparent high frequency of HJs around stars younger than a few tens of Myr (\citealt{donati16,2017MNRAS.467.1342Y}, and recently \citealt{2020AJ....160...33R}) relative to their older counterparts.
This finding places strong constraints on HJ migration timescales, showing that planet-disk interactions may play a significant role in the genesis of such planets. In this respect, one cannot neglect the role of dynamical interaction of planets with perturbing stars within a cluster as well as planet-planet interactions in a multiple planetary system \citep{cai17,flammini19,vanelteren19}. Moreover, a short migration timescale would imply that HJs could undergo strong XUV irradiation from their hosts, sufficient to remove their outer envelope and modify their physical properties with time (e.g., \citealt{locci19}). 
This knowledge still relies on a small number of discoveries, since a robust confirmation of the presence of such young planets is generally difficult. Indeed, the typically very high levels of activity of the host stars hamper detections, in particular for blind searches using the radial velocity (RV) method.  

Even when the evidence of a planetary companion is found, for instance through transits observed in the light curves of \textit{Kepler}/\textit{K2} and \textit{TESS}, the amplitude of the RV signal generated by the stellar activity could be up to several hundreds of \ms, and the planetary signal could go undetected even using sophisticated modelling to account for the activity of the host star. All this makes the measurement of planetary mass and bulk density very challenging.
Exemplary cases are represented by two of the handful of exoplanetary systems younger than $\sim$20 Myr. The first is represented by the super-Neptune sized companion to the low-mass star K2-33 ($R_{\rm p}=0.451\pm0.033$ $\Rjup$), a M3V star in Upper Scorpius \citep[age 11 Myr;][]{2016Natur.534..658D,2016AJ....152...61M}. The expected RV semi-amplitude due to the planet is about 20 $\ms$, which is dwarfed by the activity variability of the order a few hundred of $\ms$. A reliable mass determination is still lacking for this object, preventing the understanding of the planet bulk structure and further studies of its evolution at early stages based on solid observational results. The second case, still more complicated, is represented by the multi-planetary system V1298 Tau (age $\sim$20 Myr) detected by \textit{Kepler}/\textit{K2} \citep{david2019}, with a Jupiter-sized planet cohabiting with three more companions, all between the size of Neptune and Saturn. The high-amplitude variability in the RVs caused by stellar activity ($\sim$200 \ms over nearly five days, as measured from Keck/HIRES VIS RVs by \citealt{david2019b}), and dynamical effects due to mean-motion resonances, makes the characterisation of the V1298 Tau system very challenging with RV follow-up.   
More recently, the detection with \textit{TESS} and \textit{Spitzer} of a 0.4 $\Rjup$ transiting the bright, pre-main-sequence M dwarf AU Mic every $\sim$8.5 d \citep[age $\sim$20 Myr]{plavchan20} made headlines, in that the planet AU Mic\,b co-exists with a debris disk. The RV follow-up of the star revealed a variability due to stellar activity with amplitudes of $\sim$150 and 80 \ms in the visible and near-infrared, respectively, that allowed only for a measurement of the planet mass upper limit ($<$0.18 $\Mjup$, or $K<28$ \ms, at 3$\sigma$ confidence).   

In 2012, the Global Architecture of Planetary Systems (GAPS) project \citep{2013A&A...554A..28C} started a large and diversified RV campaign with the HARPS-N spectrograph \citep{cosentino14} at the \textit{Telescopio Nazionale Galileo} (TNG) focused on exoplanetary science. One main goal pursued by GAPS is assessing the planet occurrence rates around different types of stars (e.g., \citealt{2019A&A...621A.110B}), and understanding the origin of planetary-system diversity. Since 2017, the characterisation of exoplanetary atmospheres \citep{2019A&A...631A..34B,2020ApJ...894L..27P,2020arXiv200505676G} and the RV search for planetary companions around young stars \citep{carleo18,2020A&A...638A...5C} became main scientific themes. The RV survey was specifically designed to confirm the emerging evidence for a higher frequency of planets around young T Tauri stars than around more evolved stars (e.g., \citealt{2017MNRAS.467.1342Y}, underlying, however, that the sample is still too small for any reliable statistics), and to determine their orbital and physical parameters for a comparison with the older population. 

Within this framework, we monitored both a sample of targets in young associations (e.g Taurus, Cepheus, AB Doradus, Coma Berenices, Ursa Major) to search for planetary companions, and a small sample of targets with confirmed or candidate planets from other surveys (e.g., \citealt{2020A&A...638A...5C}). Among them, we observed the weak-line T-Tauri star \object{V830 Tau} (age $\sim$2 Myr), known to host a HJ (minimum mass $m_{\rm b}\sini_{\rm b}=0.57\pm$0.10 $\Mjup$; $P_{\rm b}=4.927\pm$0.008 d) announced by \cite{donati16}, and further characterised by the same team \citep[hereafter DO17]{donati17}. This detection came as a breakthrough, then followed by the transiting HJ HIP 67522\,b \citep{2020AJ....160...33R}, since it showed that giant planets can form at the very early stages of star formation and migrate within a gaseous protoplanetary disk. However, the host star V830 Tau shows a very high level of activity, with a RV scatter of $\sim$660 $\ms$ as measured by DO17, which is at least an order of magnitude higher than the semi-amplitude of the detected planetary signal. Hence, we became interested in observing this star with a different spectrograph, aiming to confirm the presence of the planet and refine the planetary orbital and physical parameters with a careful treatment of the RV activity signal, whilst also monitoring simultaneously the star with a dedicated photometric follow-up.   

In this paper we present the results of our independent 2.5-year long follow-up campaign of \object{V830 Tau}. It is structured as follows. We describe the original datasets used in our analysis in Section \ref{sec:datasets}, and present updated stellar fundamental parameters derived from HARPS-N spectra in Section \ref{sec:stellarparam}. A characterisation of stellar activity observed during our campaign is given in Section \ref{sec:stellaractivity}, where we also present the results of a transit search for \object{V830 Tau\,b} in the \textit{Kepler/K2} light curve. In Section \ref{sec:rvextraction} and \ref{sec:rvanalysis} we discuss several methods and techniques that we used to extract and model the RVs for characterising the planetary signal. We support our conclusions about our search for planet\,b by quantifying the detection limits through injection-recovery simulations (Section \ref{sec:simulations}), and present a summary and final discussion in Section \ref{sec:discussion}.   

\section{Description of the datasets}
\label{sec:datasets}

\subsection{HARPS-N spectra}

We collected 146 spectra of \object{V830 Tau} with the HARPS-N spectrograph \citep{cosentino14} between 17 October 2017 and 15 March 2020 (timespan 878~d), almost two years after the observations of DO17, with a median S/N=19.6 measured over all the \'echelle orders. We excluded from further analysis the spectra collected at epochs BJD 2458052.702 and 2458098.589, which have S/N=4.1 and 0.5, respectively.
HARPS-N is a cross-dispersed high-resolution (R=115\,000) and high-stability \'echelle spectrograph, covering the wavelength range 3830 -- 6930 \AA. The spectrograph is fed by two fibres, one on the target and the second one, used as a reference, illuminated by the sky in the case of V830 Tau.
 
Following the method described in \cite{malavolta17a}, we did not find evidence for spectra contaminated by moonlight, which in general could affect the RV of the target in a measurable way. 

During the last season, from 16$^{th}$ Nov 2019 to 28$^{th}$ Feb 2020, we adopted a denser sampling by scheduling the target twice per night whenever possible, with the pair of observations separated by at least three hours. This change in the observing strategy was intended to improve the fit of the short-period component of the activity signal, modulated over the known $\sim$2.7-d rotation period, especially during consecutive nights of observation.

\subsection{Photometric light curves}
\label{subsec:k2stelladata}

\object{V830 Tau} was observed by the \textit{Kepler} extended \textit{K2} mission in long-cadence mode (one point every 30 minutes) during campaign 13, from 2017-03-08 to 2017-05-27 (K2 target ID EPIC 247822311). 
Almost 1.5 years after \textit{K2} observations, we followed-up V830 Tau from Oct 2018 to the end of January 2020 with the STELLA facility in Tenerife \citep{strass04} and its wide-field imager WiFSIP. This time span corresponds to the last two seasons of our monitoring with HARPS-N. Blocks of 5 exposures per filter were collected, with exposure times per single image of 60s (\textit{V}-band) and 25s (\textit{I}-band). Standard data reduction, including bias subtraction and flat field correction, was performed, and aperture photometry used to extract the differential light curve, as described in \cite{mallonn18}. We then analysed the averaged values per observing block and filter, resulting in 125 and 122 data points for the \textit{V} and \textit{I} filter, respectively. The ensemble of comparison stars (UCAC4 573-011610, 573-011607, 574-011176, 573-011630) was chosen in automatic by the pipeline to minimise the scatter of the differential photometry. However, due to the strong variability of the target, a specific choice of the reference stars has minor effects on the differential light curve. The data are listed in Tables~\ref{table:stellavband} and \ref{table:stellaiband}.


\section{Stellar parameters and lithium abundance}
\label{sec:stellarparam}

Spectroscopic determination of stellar parameters via standard techniques based on line equivalent widths (EWs) and imposing excitation/ionisation equilibria is not possible for our star because of its low effective temperature ($T_{\rm eff}$) and relatively rapid rotation. We therefore performed a spectral synthesis analysis of the co-added spectrum of the target in four spectral regions around $\lambda\lambda$5400, 5800, 6200, 6700\,\AA, and using the 2017 version of the MOOG code (\citeyear{sneden73}; 2017 version) with the driver {\it synth}. We considered the line list kindly provided by Chris Sneden (priv. comm.), the \cite{castelli04} grid of model atmospheres, with solar-scaled chemical composition and new opacities ({\sc ODFNEW}), and linear limb-darkening coefficient taken from \cite{claret2019}. Assuming [Fe/H]=0.0 dex as iron abundance, in compliance with the metallicity of the Taurus-Auriga star forming region as derived by \cite{dorazi11}, microturbulence velocity of\,1.0\,km\,s$^{-1}$ and macroturbulence of 1.4\,km\,s$^{-1}$ by \cite{brewer2016}, we varied the effective temperature in the range of 3800-4500\,K and surface gravity in the range 3.6-4.5\,dex at steps of 100\,K and 0.1\,dex, respectively, until the best match between target and synthetic spectra was found for each spectral region. In the end, the mean values of effective temperature obtained from the four spectral regions were adopted. Final values of the derived parameters were $T_{\rm eff}=4050\pm100$\,K and $\log g=3.95\pm0.20$\,dex, where the errors take into account both the best fit determination and the standard deviation of the mean obtained from the four spectral regions considered. 
Our estimate for $T_{\rm eff}$ is slightly lower than that reported by DO17, who adopted $T_{\rm eff}=4250\pm50$ K from \cite{donati2015}, but consistent with values found by other authors (see, e.g., \citealt{sestitoetal2008}, and references therein). As a by-product, we also derived a projected rotational velocity of $v\sin i= 30\pm1$\,km\,s$^{-1}$, consistent with the value 32.0$\pm$1.5 \kms reported by \cite{nguyenetal2012}. 

Once stellar parameters were derived, we also measured the EW for the lithium doublet at 6708 \AA, finding $EW$(\ion{Li}{i})=658$\pm$5 m$\AA$, which corresponds to an abundance $A$(\ion{Li}{i})=3.19$\pm$0.16 dex (\citealt{lind09}). Note that the error on $T_{\rm eff}$ is the dominant source of uncertainty, implying an error of 0.15 dex in $A$(\ion{Li}{i}). This means that our measured Li abundance is in agreement, within the uncertainties, with values measured in T Tauri stars and meteoritic abundance. Furthermore, our value is consistent, within the errors, with previous findings using the same method (see \citealt{sestitoetal2008}, and references therein). Applying non-LTE (NLTE) corrections and following the prescription by \cite{lind09}, we obtain $A$(\ion{Li}{i})$_{\rm NLTE}=3.16$~dex. 

\section{Stellar activity analysis}
\label{sec:stellaractivity}
\subsection{Light curve analysis}
\label{subsec:photometryanalysis}

\begin{figure}
   \centering
   \includegraphics[width=\hsize]{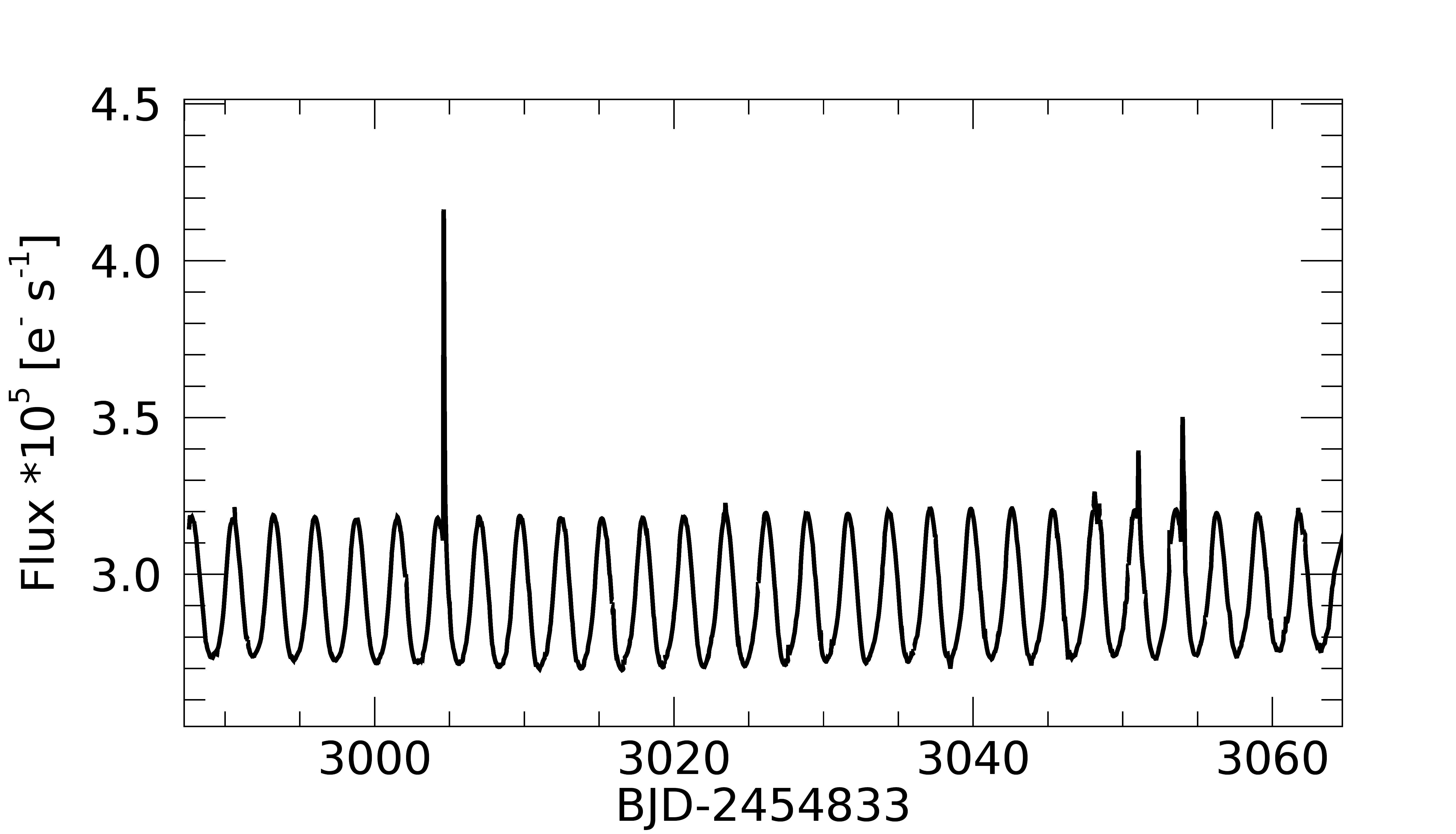}\\
   \includegraphics[width=0.95\hsize]{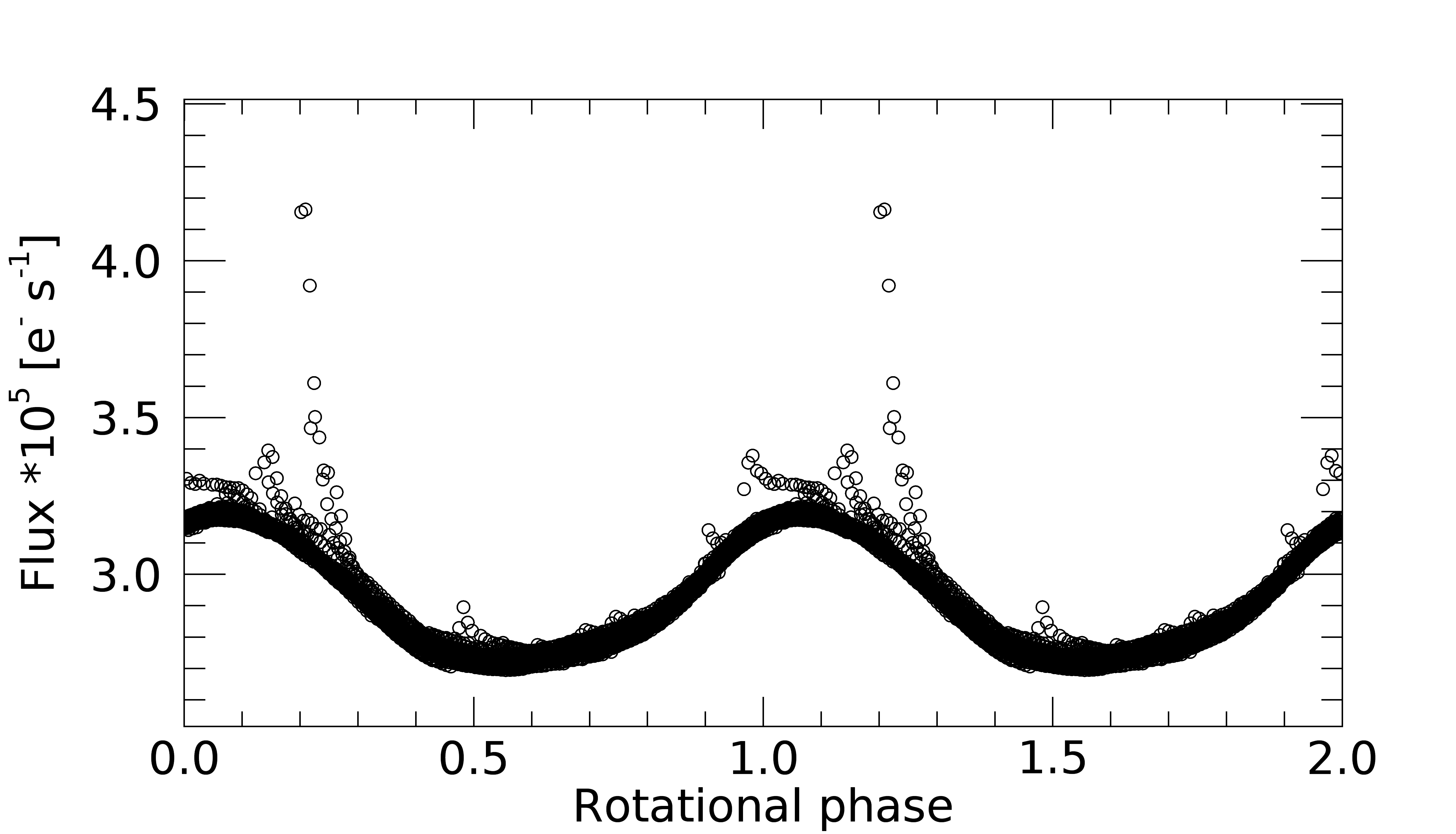}\\
  \caption{\textit{Kepler/K2} light curve. The time series collected during the first half of 2017, and the same data phase-folded to the stellar rotation period are shown in the upper and lower panel, respectively. The most powerful flares occurred at phases close to the maximum brightness.}
\label{Fig:lck2}
\end{figure}

\begin{figure*}
   \centering
   \includegraphics[width=\hsize]{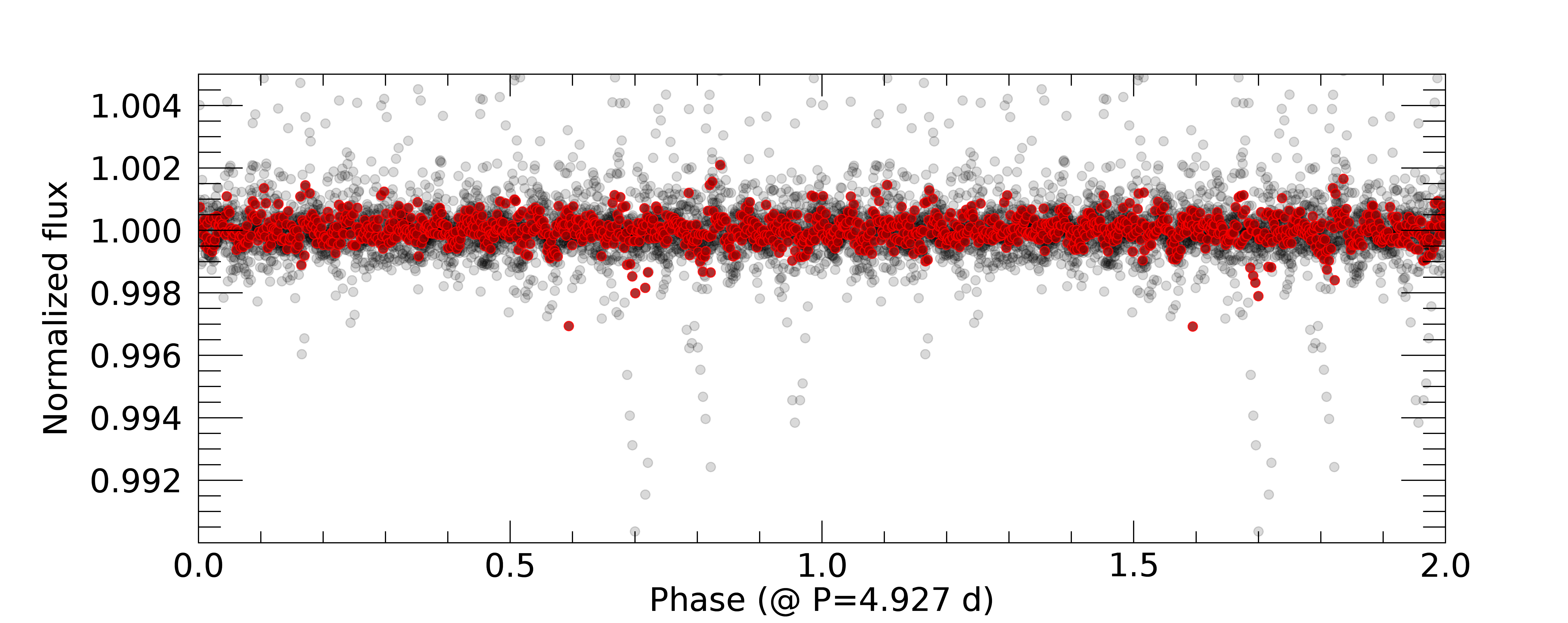}\\
  \caption{Detrended and flattened \textit{K2} light curve of \object{V830 Tau}, phase-folded at the orbital period of the planet announced by DO17. Red dots represent 5-point binned data.}
         \label{Fig:lk2foldedplanet}
\end{figure*}  

\begin{figure*}
   \centering
   \includegraphics[width=\hsize]{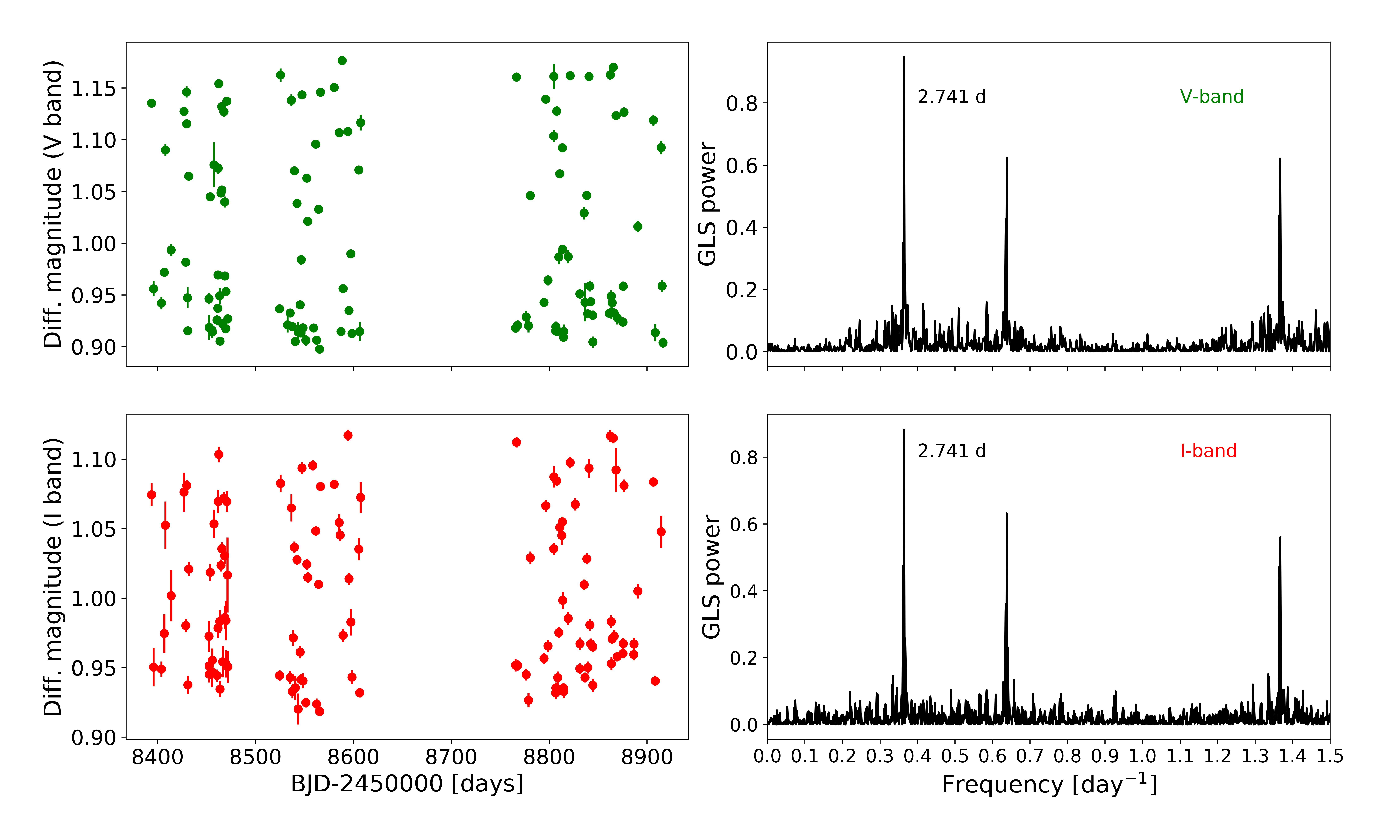}\\
  \caption{STELLA light curves in $V$- and $I$-bands, spanning the last two seasons of HARPS-N observations, and their respective GLS periodograms. The peak corresponding to the rotation period has been labelled. }
         \label{Fig:lcstella}
\end{figure*}   

Figure~\ref{Fig:lck2} shows the time series of \textit{Kepler/K2} photometry and the data phase-folded to the known rotation period $P_{\rm rot}=2.74$~d. 
We corrected the K2 light curve following the procedure described in detail in \citet{Nardiello2016} and \citet{Libralato2016}. Briefly, we performed a two-step correction: first, we corrected the systematic trends associated with the conditions of the spacecraft, of the detector and the environment. In order to perform this correction, we fitted and applied to the light curve of the target a linear combination of orthonormal bases, the co-trending basis vectors, released by the Kepler team\footnote{\url{https://archive.stsci.edu/k2/cbv.html}}. In a second step, we corrected the  position-dependent systematics due to the large jitter of the spacecraft. We refer the reader to Sect.~3.1 of \citet{Nardiello2016} for a detailed description of the corrections. 
We flattened the light curve as done in \citet{Nardiello2019,Nardiello2020}: we defined a number of knots spaced 6.5h on the light curve, and interpolated these knots with a 5th order spline to obtain a model of the stellar variability. We used this model to flatten the light curve. In this process, we also excluded all the points 3$\sigma$ above the average value of the flattened light curve.
The corrected photometric data, phase-folded at the orbital period $P=4.927$~d found by DO17, is shown in Fig.~\ref{Fig:lk2foldedplanet}. We did not detect any transit signal at the period of the claimed planet, as further confirmed by the analysis with the Box Least-Squares algorithm (BLS; \citealt{kovacs02}). For comparison, assuming $R_{\star}$=2 $\Rsun$ (after DO17), and a planetary radius between 1.5 and 2 $\Rjup$, estimated from the giant planet thermal evolution models by \cite{fortney07} for a planet of mass, age and orbital distance (scaled to take into account the irradiation of an M star, according to its luminosity) as those found by DO17, we would expect a transit with depth between $\sim$0.6 and 1$\%$, i.e. 6--10 times the RMS of the flattened light curve.

Figure~\ref{Fig:lcstella} shows the differential photometry of STELLA and their Generalized Lomb-Scargle periodograms (GLS; \citealt{zech09}), that show clear peaks at the rotation period and 1-d aliases. We note that the full amplitude of the STELLA $V$-band photometry is the same ($\Delta V$=0.28 mag) as that measured by DO17 between 2015 Oct 30 and 2016 Mar 15.

\begin{figure*}
   \centering
  \includegraphics[width=\hsize]{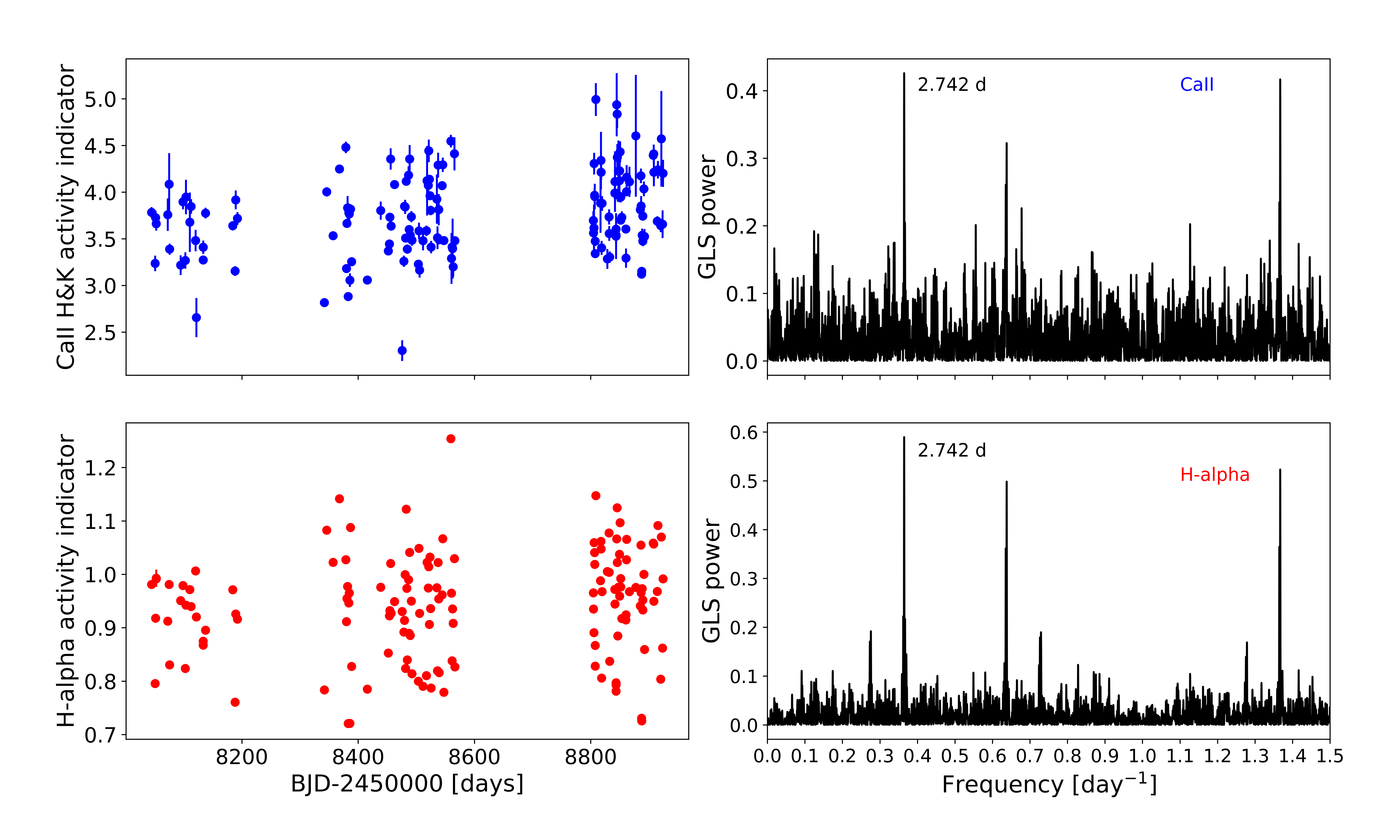}
  \caption{ Time series of the activity indicators extracted from the HARPS-N spectra based on the CaII H$\&$K and H-$\alpha$ spectral lines (left panels), and their corresponding GLS periodograms (right panels). The peak corresponding to the rotation period has been labelled. }
  \label{Fig:caII_halpha_actind}
\end{figure*}

\subsection{Spectroscopic diagnostics}
\label{subsec:actspecdiagnostic}

We extracted the chromospheric activity indexes based on the CaII H$\&$K and H$\alpha$ spectral lines using the method described in \cite{gomesdasilva11} and the code \texttt{ACTIN v1.2.2} \citep{gomesdasilva18}.
Their time series and GLS periodograms are shown in Fig.~\ref{Fig:caII_halpha_actind}. Peaks at the rotation period and 1-d aliases are clearly visible for both indicators, which show an increase in the level of the stellar magnetic activity over the time span of our spectroscopic follow-up.

\begin{figure*}
   \centering
  \includegraphics[width=0.45\hsize]{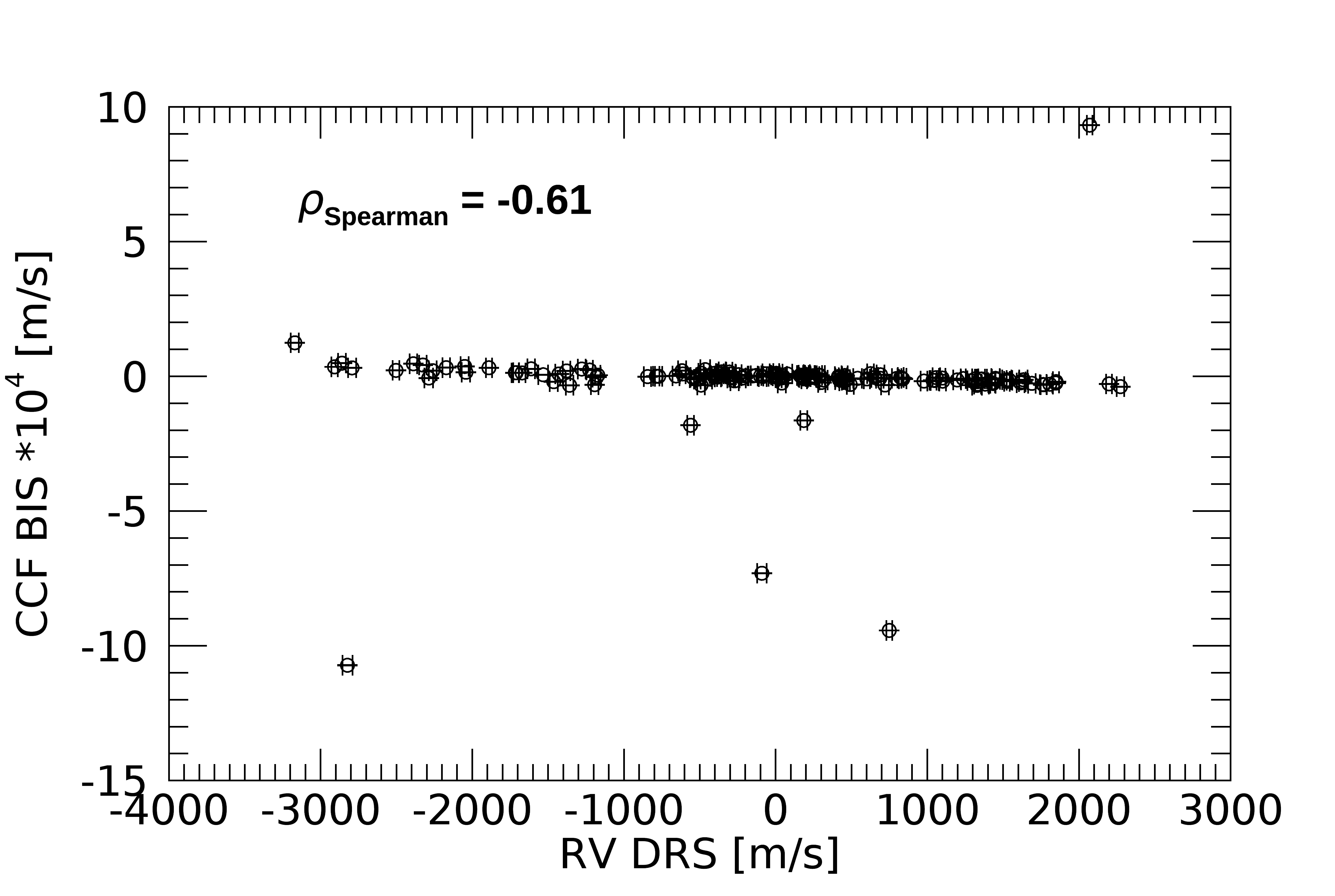}
  \includegraphics[width=0.45\hsize]{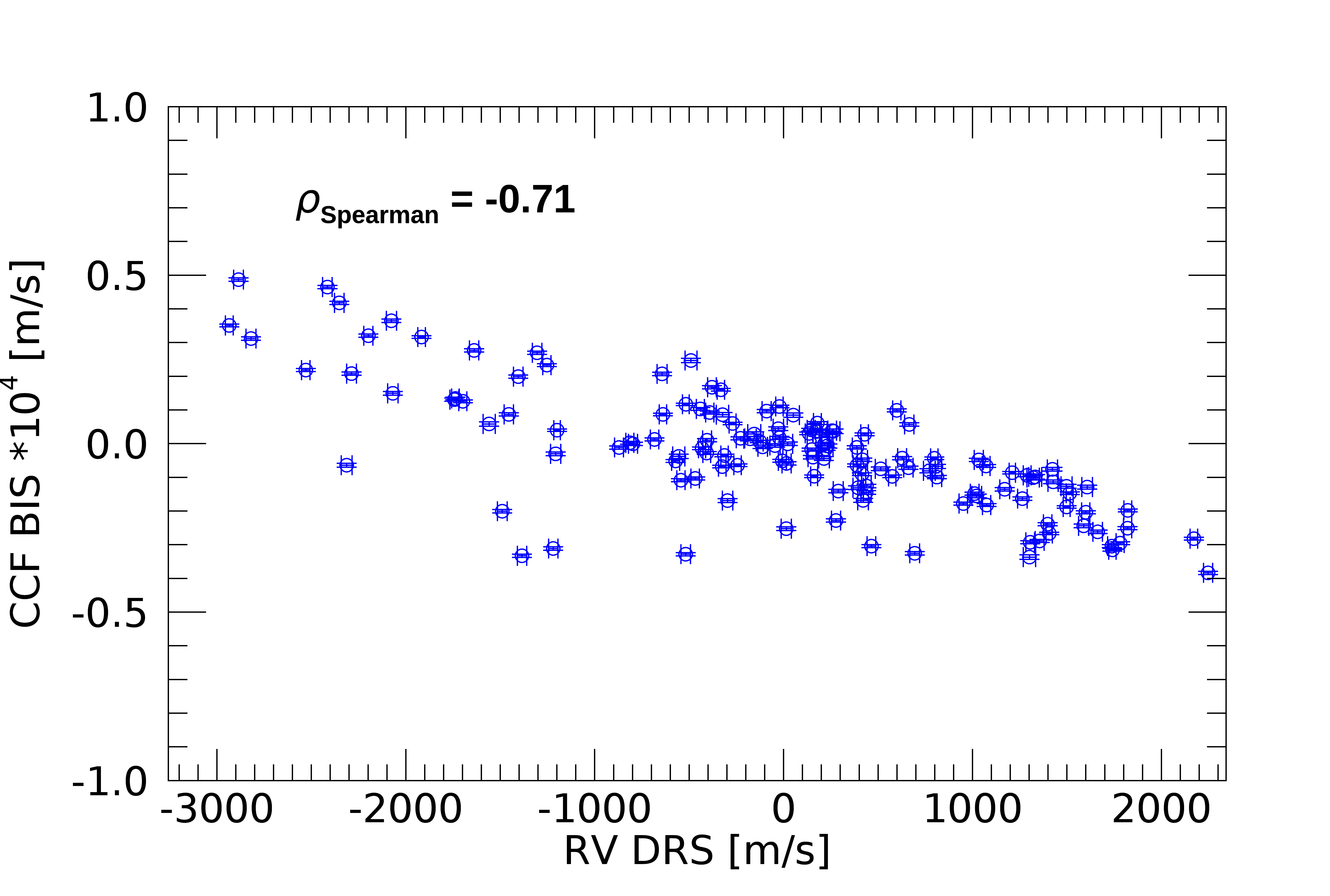}
  \caption{Correlation plot between the CCF BIS index and the radial velocities extracted with the DRS pipeline (see Sect.~\ref{sec:rvextraction}). Left panel: all data. Right panel: data without outliers, for a better visualisation of the RV-BIS correlation.}
  \label{Fig:rvbiscorr}
\end{figure*}

As we will show in Sect.~\ref{sec:rvextraction}, the cross-correlation function (CCF) of the HARPS-N spectra appears highly distorted due to the high level of stellar activity. Therefore, activity indicators based on a measure of the CCF asymmetry, such as the full width at half-maximum (FWHM) and bisector inverse slope (BIS) could be unsuitable to correct for the activity signal in the RVs. This is evident from Fig.~\ref{Fig:rvbiscorr}, showing the correlation between the BIS index and one of the RV datasets considered in this study, and where seven BIS outliers are visible. With this caveat, some other indicators of the CCF shape were further considered in our RV modeling attempts, as discussed in Sect.~\ref{subsec:kernreg}.
 
\begin{figure}[h]
\centering
\includegraphics[width=\linewidth]{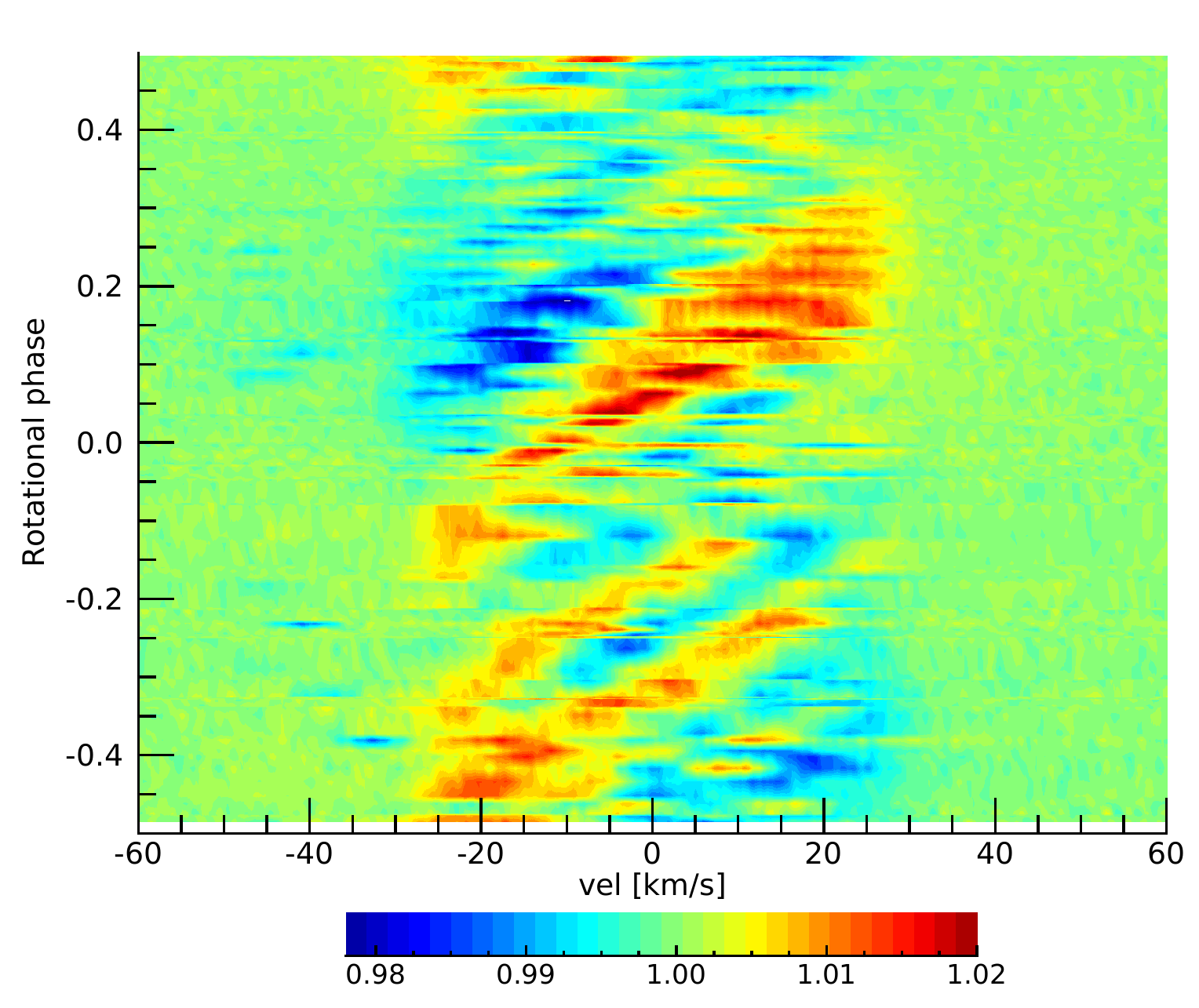}
\caption{Residuals of the cross-correlation function (CCF) -i.e.\ each individual CCF is divided by the average CCF calculated over the whole dataset- phase folded to the rotational period of V830 Tau (2.741 d).}
\label{fig:CCFresiduals}
\end{figure}

We also analysed the stellar activity by phasing the CCF residuals to the rotational period, i.e.\ the CCFs divided by the average CCF calculated over the whole dataset (Fig.~\ref{fig:CCFresiduals}).
While the shape of the map qualitatively confirms the rotational period of the star of 2.741~d, it is interesting to note that active regions
(positive and negative deviations from the average CCF) are moderately stable in position over the three seasons of observations. 

The time series of the spectroscopic activity diagnostics are listed in Table~\ref{table:actinddata}.
\section{Extraction of the radial velocities with different methods}
\label{sec:rvextraction}

\begin{figure}
   \centering
   \includegraphics[width=\hsize]{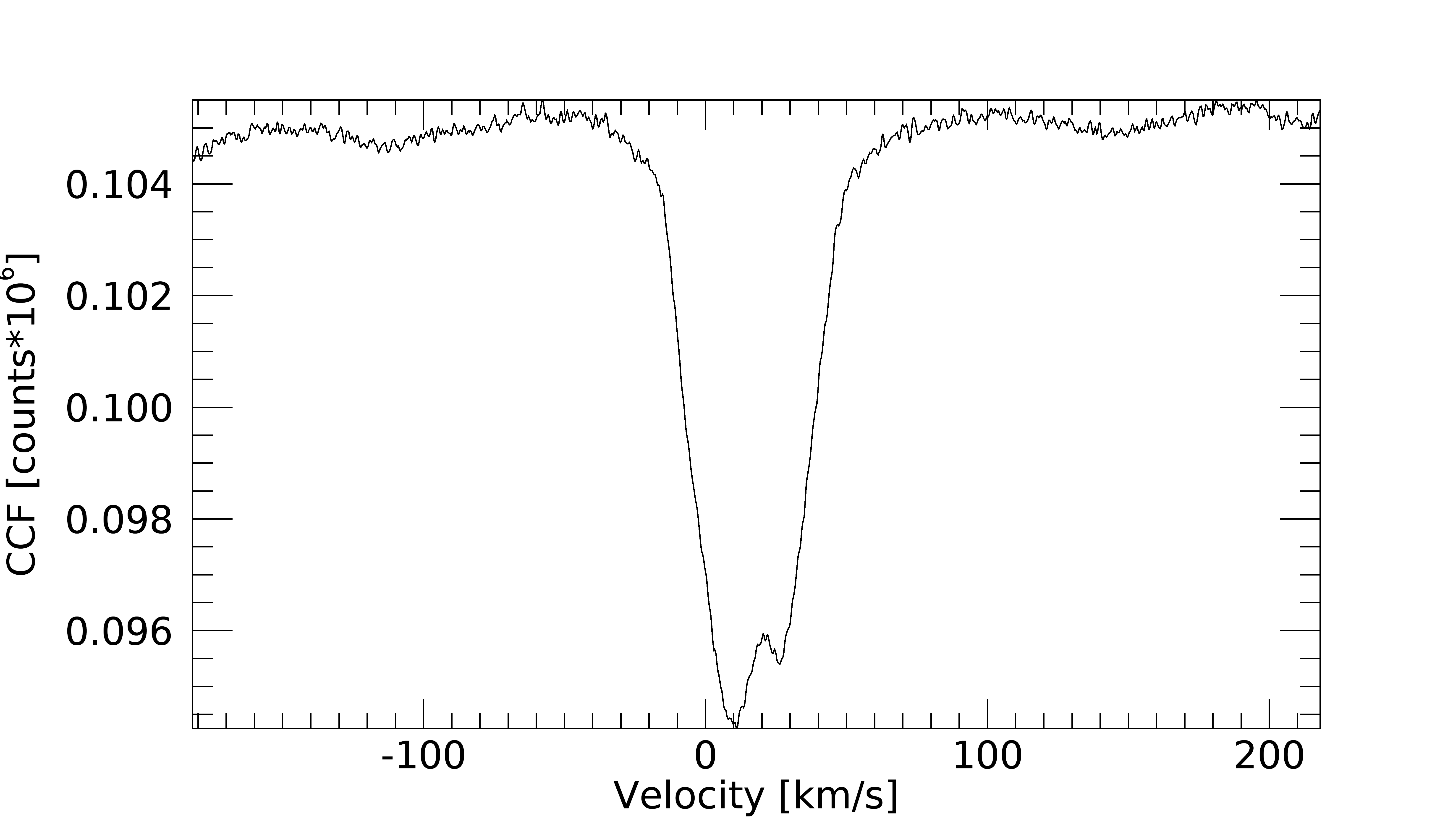}
      \caption{Cross-correlation function (CCF) of V830 Tau for a HARPS-N spectrum with average S/N. The CCF was calculated by the DRS pipeline using a template mask for a KV5 star. 
              }
         \label{Fig:ccfprofile}
\end{figure}
To deal with the very challenging task of detecting a signal induced by the Keplerian motion of V830 Tau b, whose expected semi-amplitude is more than an order of magnitude smaller than the RV scatter due to magnetic activity, we extracted RVs using three independent methods, and different from the least-squares deconvolution (LSD) used by DO17. Each extraction method could be affected by the very high stellar activity in a different way, this is why we decided to analyse different datasets in the attempt of detecting and characterising the signal induced by \object{V830 Tau}\,b.

We used the standard DRS pipeline version 3.7.0 to extract the RVs through the CCF technique \citep{2002A&A...388..632P}. To calculate the CCF we adopted the template mask for a K5V dwarf and a half-window of the CCF of 200 $\kms$, to account for the line broadening due to the fast stellar rotation and to include a good portion of the continuum for a proper fitting of the CCF profile. Figure \ref{Fig:ccfprofile} shows the CCF of one HARPS-N spectrum with average S/N, which illustrates clearly the strong deformation in the core of the average line profile due to stellar activity. 

We also used the template-matching TERRA pipeline \citep{anglada12} to extract an independent dataset. Our default TERRA dataset is that obtained by considering orders corresponding to a reference wavelength higher than $\lambda=4530$\,\AA , as recommended for stars with a spectral type like that of \object{V830 Tau}. The computation of the RVs includes a correction for the secular perspective acceleration.

Finally, we extracted the RVs from the observed spectra using the Gaussian process-based, template-free approach recently proposed by \cite{rajpaul20} (hereafter identified as the dataset \textit{R20}). In brief, Gaussian processes (GP) are used to model and align all pairs of spectra with each other; the pairwise RVs thus obtained are combined to produce accurate differential stellar RVs, without having to construct a template. Such differential RVs can be extracted on a localised basis, for example\ to yield an independent set of RVs for each \'echelle order, or indeed for much smaller sub-divisions of orders. The rationale behind the latter approach is that regions of spectra affected by, for instance, stellar activity or telluric contribution may in principle be identified and excluded (effectively a data-driven masking of the spectrum, without any knowledge of line locations or properties) from the calculation of the final RVs, which are obtained via an inverse variance-weighted average of the localised RVs.
The RVs used in this work were obtained by combining RVs from each \'echelle order, and without any masking. This approach was found to yield the highest signal to (estimated) noise ratio. We did, however, explore several alternative schemes for RV extraction, where we divided each order into anything from $16$ to $128$ ``chunks'', each of which might have contained between zero and several lines, and then selectively re-combined these localised RVs trying to produce a final set of RVs that e.g.\ minimised correlations with the FWHM or BIS time series, minimised periodogram power near the stellar rotation period of $2.74$~d, and/or maximised the power near the putative planetary orbital period of $4.93$~d. We explored both iterative optimisation schemes and more sophisticated machine learning approaches (e.g.\ the HDBSCAN algorithm; \citealt{campello13}) for optimising the masking procedure. Unfortunately, we found that virtually all the useful Doppler information was contained in spectral regions strongly contaminated by rotational activity: all attempts to suppress this stellar signal while trying to boost periodicity at $4.93$~d led to RV error bars that were at least an order of magnitude larger ($>250 \ms$) than in the mask-free case, thus thwarting our attempts to ``tease out'' the putative planetary signal.

\begin{figure}
   \centering
   \includegraphics[width=\hsize]{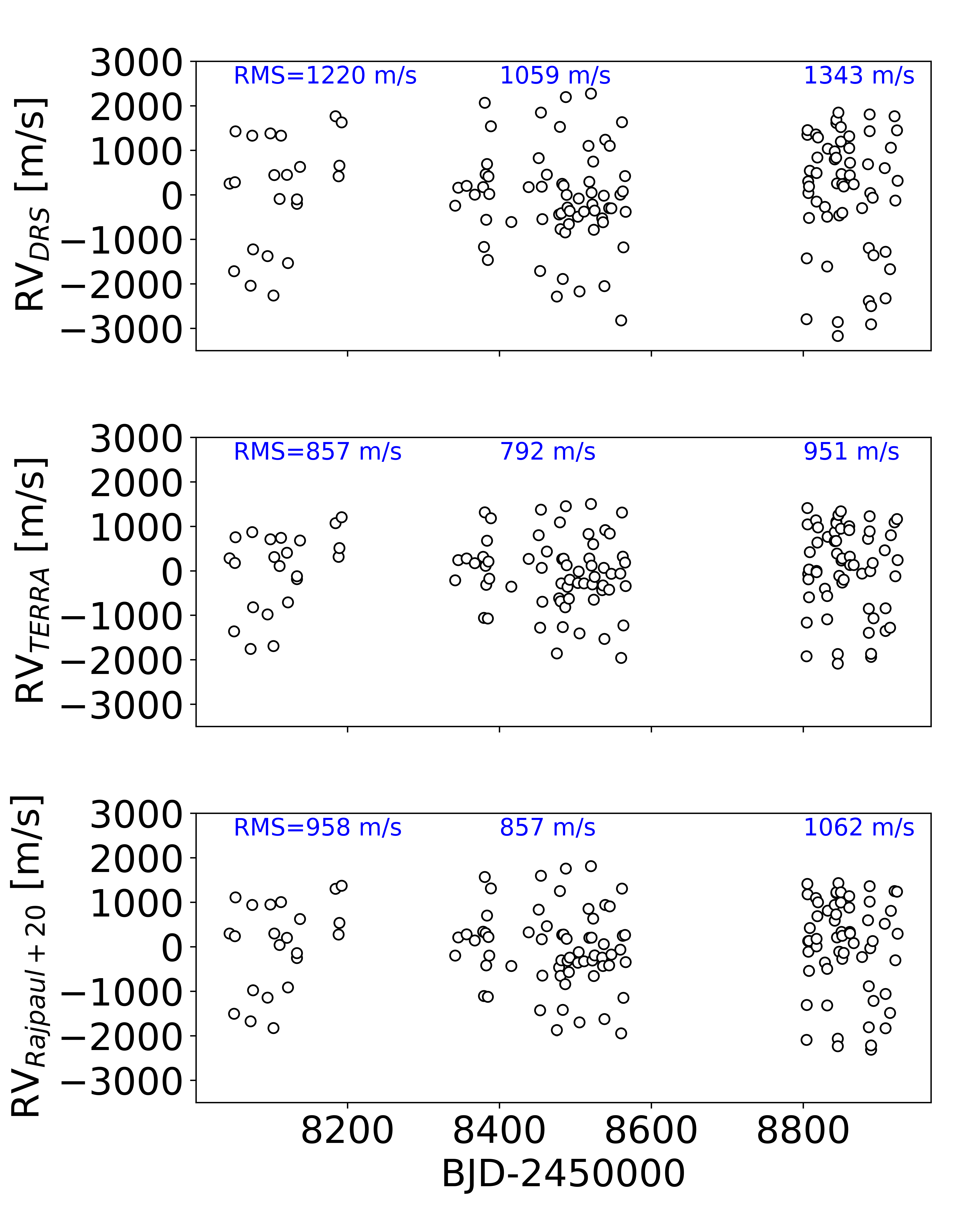}\\
    \caption{Radial velocities of \object{V830 Tau} extracted from the HARPS-N spectra with different methods (average subtracted). From top to bottom: DRS, TERRA, and template-free algorithm (R20) by \cite{rajpaul20}. The error bars are nearly two orders of magnitude smaller than the major ticks on the y-axis, and  are not visible. For each season we indicate the RMS of the corresponding RV subsample.}
    \label{Fig:rvtimeseries}
\end{figure}

We list in Table~\ref{table:rvdata} the datasets used in this work, and show the time series in Fig.~\ref{Fig:rvtimeseries}. We note that the third season is characterised by a higher RV dispersion, likely due to the increasing stellar activity visible in the spectroscopic diagnostics, as discussed in Sect.~\ref{sec:stellaractivity}. We summarise in Table~\ref{table:rvproperties} the main properties of each dataset compared to that of the MaTYSSE large programme RVs analysed by DO17, that were collected over 91 days with the ESPaDOnS and Narval spectropolarimeters linked to the 3.6-m Canada–France–Hawaii,
the 2-m Bernard Lyot, and the 8-m Gemini-North telescopes. ESPaDOnS and Narval collect spectra covering the wavelength range 3700--10000 \AA, that overlaps with that of HARPS-N and extends to the NIR region, with
a resolving power of 65\,000.

We note that the median internal error $\sigma_{\rm RV}$ of the HARPS-N RVs (TERRA and DRS extraction) is nearly half that of the MaTYSSE data, and that the TERRA dataset has a scatter reduced by $\sim$28$\%$ and 9$\%$ with respect to that of the DRS and R20 extractions, respectively. 
\begin{table*}
  \caption{Properties of the radial velocity time series extracted with TERRA, DRS, and the pipeline R20 by \cite{rajpaul20}, analysed in this work.}          
  \label{table:rvproperties}      
  \centering                 
  \begin{tabular}{lccccc}       
  \hline\hline              
  \textbf{Dataset and RV extraction method} & \textbf{Time span} & \textbf{No. RVs} & \textbf{RV RMS} & \textbf{Median $\sigma_{\rm RV}$} & \textbf{$\sigma_{\rm RV}$ RMS} \\    
  & [days] & & [$\ms$] & [$\ms$]  & [$\ms$] \\ 
  \hline                        
  \textit{HARPS-N}\\
  \hline
  TERRA (starting from \'echelle order no. 22) & 880 & 144 & 875 & 24 & 12 \\        
  DRS & 880 & 144 & 1213 & 24 & 4 \\ 
  R20 & 880 & 144 & 965 & 54 & 17\\
  \hline
  \textit{ESPaDOnS, Narval, and ESPaDOnS/GRACES}\\
  \hline
 Least-square deconvolution & 91 & 75 & 662\tablefootmark{a} & 51 & 10.8 \\
  \hline     
  \end{tabular}
  \tablefoot{
  \tablefoottext{a}{As published by \citet{donati17}, without any instrumental offset applied. These measurements were collected between late 2015 to February 2016}
  }
\end{table*}

\section{Radial velocity analysis}
\label{sec:rvanalysis}

We describe hereafter the results obtained from the analysis of our HARPS-N RVs. We start with showing the GLS periodograms, to illustrate that the time series are clearly dominated by signals produced by stellar activity and $\pm$1~d$^{-1}$ aliases. 

\subsection{Frequency content analysis}
\label{subsec:rvfreqcontent}

\begin{figure*}
    \begin{subfigure}{.5\textwidth}
   \centering
   \includegraphics[width=0.86\linewidth]{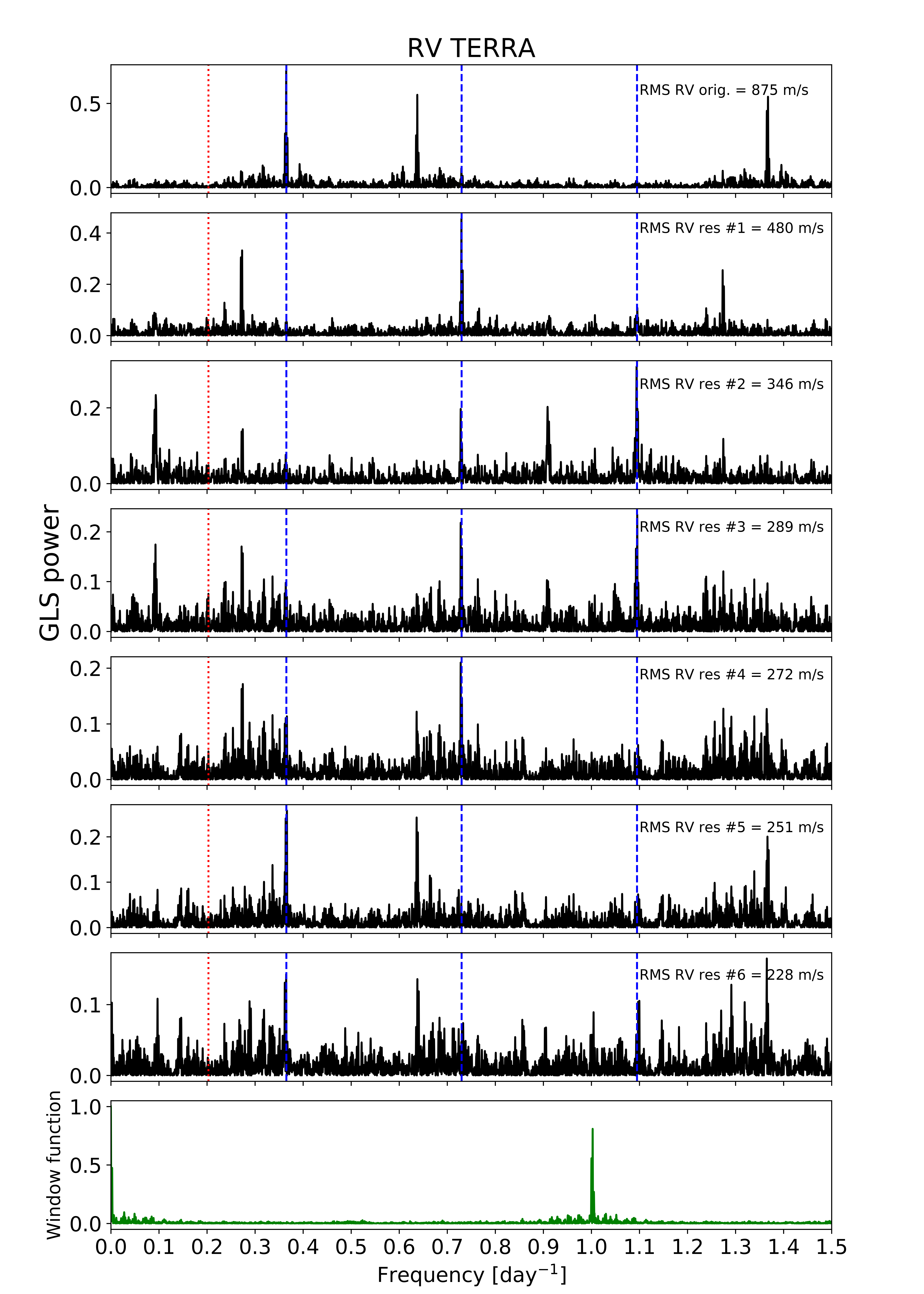}
      \caption{}
       \end{subfigure}
       \begin{subfigure}{.5\textwidth}
   \centering
   \includegraphics[width=0.86\linewidth]{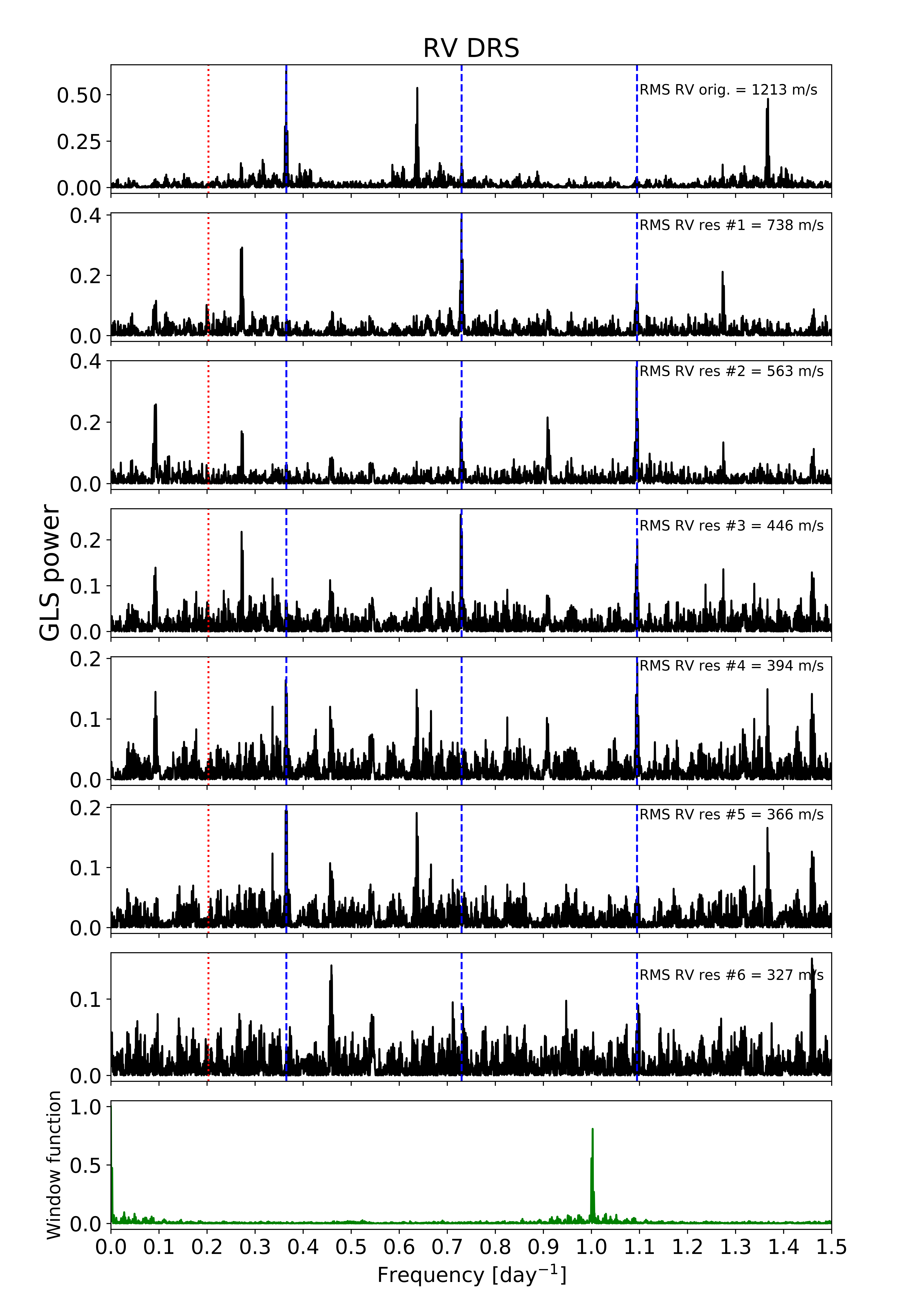}
   \caption{}
    \end{subfigure}
  \begin{subfigure}{.5\textwidth}
   \centering
   \includegraphics[width=0.86\linewidth]{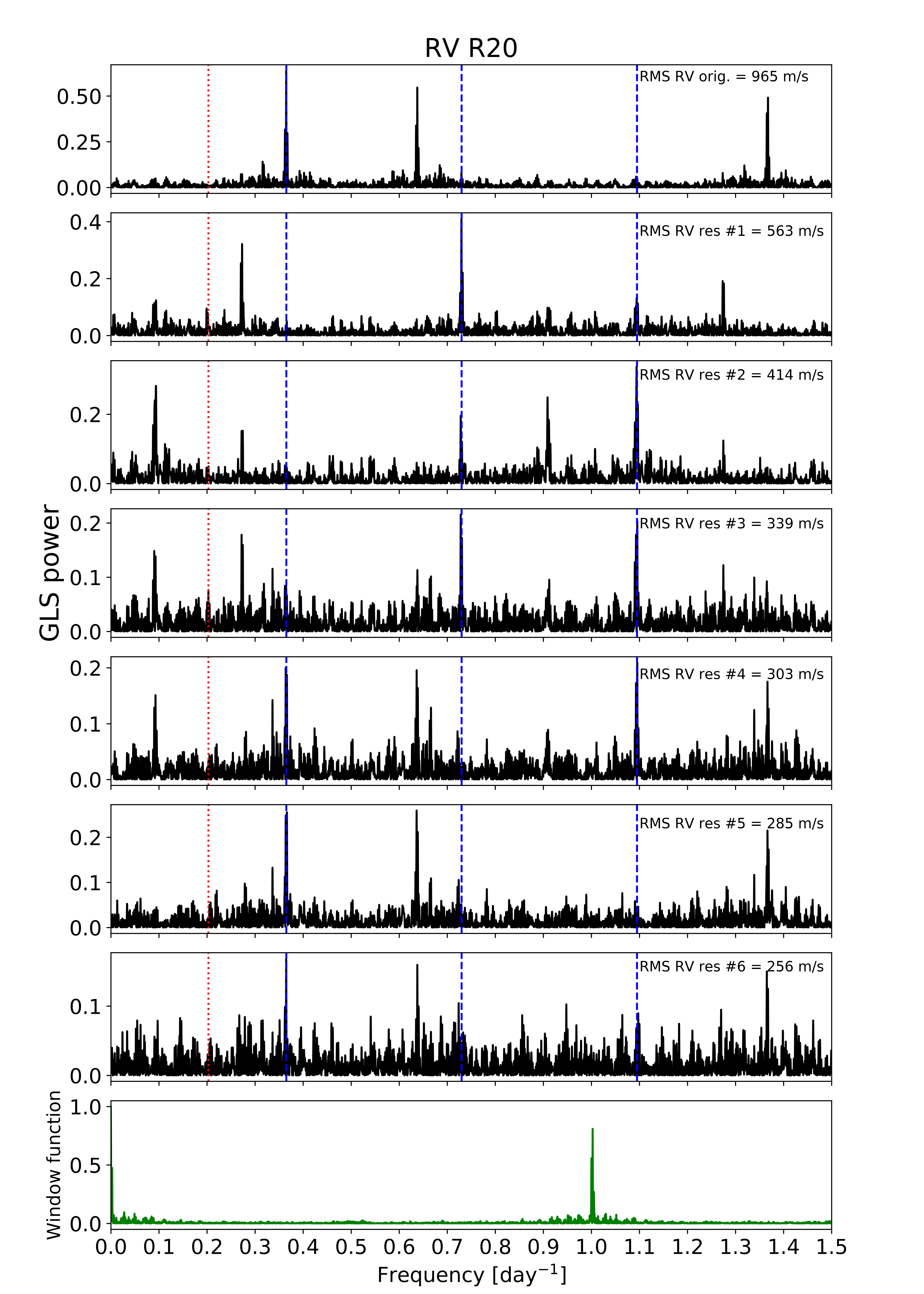} 
   \caption{}
  \end{subfigure}
    \caption{GLS periodograms of the original TERRA (fig. a), DRS (fig. b), and R20 (fig. c) RVs and their residuals after recursive pre-whithening. For each figure: the vertical and dashed blue lines indicate the stellar rotation frequency and its first and second harmonic; the red line marks the orbital frequency of the planet announced by \cite{donati16,donati17}; each panel reports the RMS of the dataset used for calculating a specific periodogram; the window function is shown in green in the bottom panel.}
    \label{Fig:glsrecursiveterra}
\end{figure*}

\begin{figure}
   \centering
   \includegraphics[width=0.8\hsize]{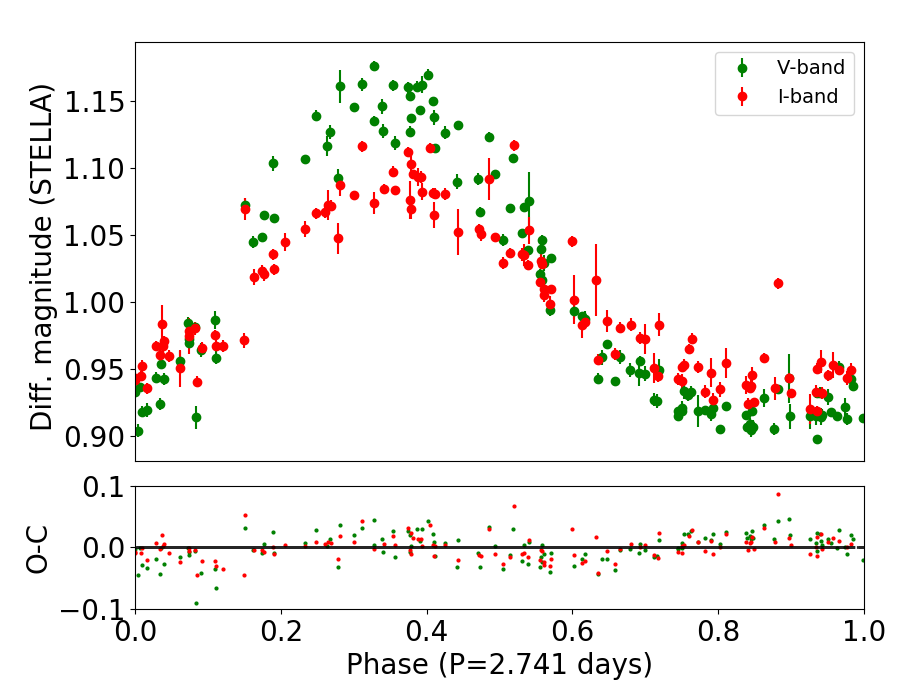}\\
   \hspace*{1.5mm}
   \includegraphics[width=0.78\hsize]{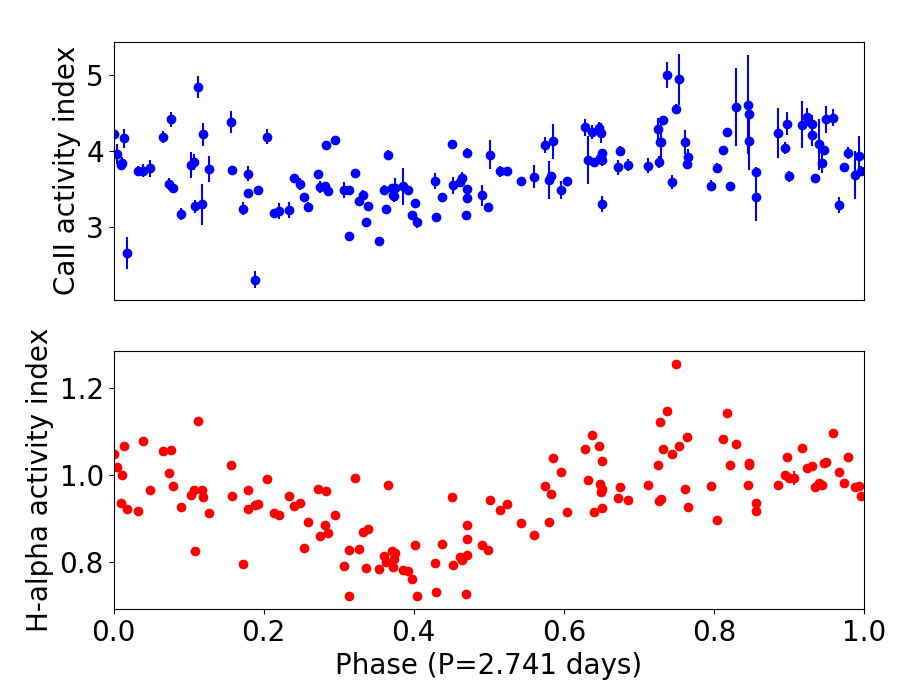}\\
   \hspace*{-4.5mm} 
   \includegraphics[width=0.86\hsize]{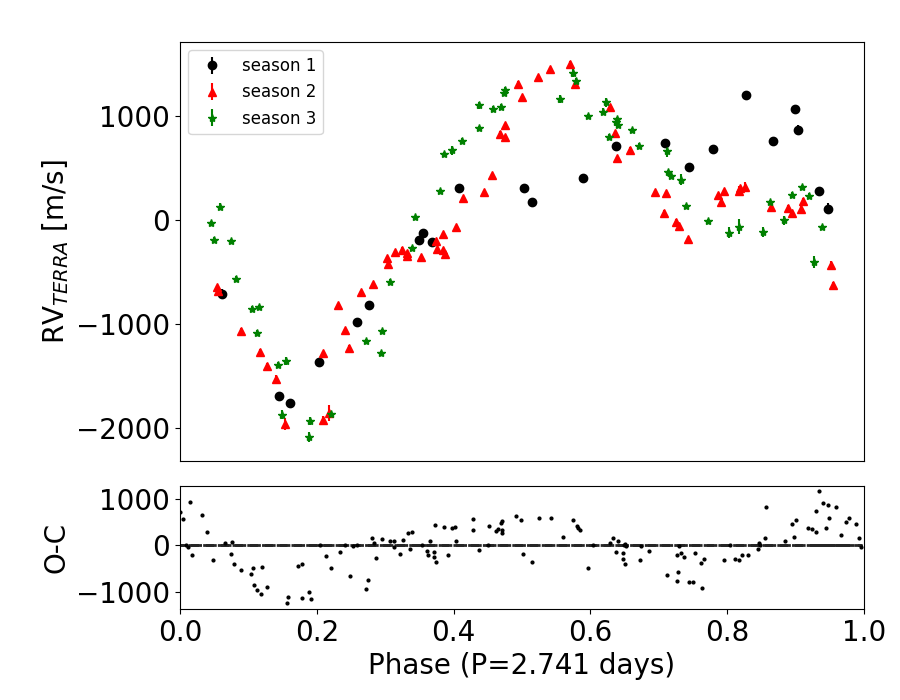}
      \caption{STELLA light curves (upper panel), CaII H$\&$K and H-$\alpha$ activity indicators (middle panel), and TERRA RVs (lower panel) phase folded using the same period $P=2.741$~d and reference epoch of the STELLA V-band photometry and\textit{ Kepler/K2} light curve in Fig. \ref{Fig:lck2}. Subplots for the light curves and RVs show the residuals $O-C$ of the best-fit sine function obtained with the GLS software.
              }
         \label{Fig:stellaterrafold}
   \end{figure}

We calculated the GLS periodograms for the original data and residuals after recursive pre-whitening, as shown in Fig.~\ref{Fig:glsrecursiveterra} for all the datasets. The periodogram of the original data (first panel) shows a very sharp maximum at the stellar rotation frequency, and signals related to stellar activity (at the rotation frequency or its harmonics, and 1-d aliases) clearly dominate the periodograms even after five pre-whitening iterations. The periodogram of the RV residuals after the last pre-whitening is still characterized by high dispersion (228 and 327 \ms\ for TERRA and DRS data respectively), around $\sim$3-5 times the semi-amplitude of the claimed signal induced by the planet. In general, pre-whitening using only sinusoids is \emph{not} an optimal way to account for stellar activity and search for planetary signals, and more sophisticated functions should be used to model the complex activity- related signals; however, these results do at least illustrate well that unearthing a small planetary signal is not a trivial task for \object{V830 Tau}. 
Taking advantage of the fact that the STELLA data span the last two seasons of the RV follow-up, it is interesting to compare the photometric and spectroscopic datasets by phase folding the data to a common period and phase-zero epoch. This can provide some insights into the nature of stellar surface patterns responsible for the periodic modulation observed in the RVs. We present this comparison in Fig.~\ref{Fig:stellaterrafold}, using the TERRA dataset. The light curves and the spectroscopic activity indicators are anti-correlated, and this can be explained by a spot-dominated activity. Such an evidence is also confirmed by noting that the \textit{I}-band light curve has a smaller amplitude. This indicates that active regions are dominated by cooler features (starspots) rather than hotter, facular-like, features. Also the $\sim$0.2 phase shift between RVs and light curves is typical of the effect related to the flux deficit due to spots that affect the CCF. We note some differences among the RV folded curves distinguished by observing season.  

\subsection{Gaussian process regression}
\label{subsec:gprvanalysis}

We turned to more sophisticated analysis techniques to mitigate the stellar activity contribution to the variability observed in the RVs. Gaussian process regression, which has been often applied to detect and characterise planetary signals in RV time series (e.g. \citealt{haywood14, dumusque17}), was used by DO17 to derive their planet parameters from the raw ESPaDOnS/Narval/GRACES RVs, and proved to be a very efficient way to model the activity over the shorter time span of their observations ($\sim$3 months). We apply here the same technique and model, using a quasi-periodic covariance matrix for the correlated signal due to stellar activity: 
\begin{eqnarray} \label{eq:eqgpkernel}
k(t, t^{\prime}) = h^2\cdot\exp\large[-\frac{(t-t^{\prime})^2}{2\lambda^2} - \frac{sin^{2}(\pi(t-t^{\prime})/\theta)}{2w^2}\large] + \nonumber \\
+\, (\sigma^{2}_{\rm RV}(t)+\sigma^{2}_{\rm jit})\cdot\delta_{t, t^{\prime}}
\end{eqnarray}
where $t$ and $t^{\prime}$ represent two different epochs, $\sigma_{\rm RV}$ is the radial velocity uncertainty, and $\delta_{t, t^{\prime}}$ is the Kronecker delta. Our analysis takes into account other sources of uncorrelated noise -- instrumental and/or astrophysical -- by including a constant jitter term $\sigma_{\rm jit}$ which is added in quadrature to the formal uncertainties $\sigma_{\rm RV}$. $h$, $\lambda$, $\theta$ and $w$ are the GP hyper-parameters: $\theta$ represents the periodic time-scale of the modelled signal, and corresponds to the stellar rotation period; $h$ denotes the scale amplitude of the correlated signal; $w$ describes the level of high-frequency variation within a complete stellar rotation; and $\lambda$ represents the decay timescale in days of the correlations, which relates to the temporal evolution of the magnetically active regions responsible for the correlated signal observed in the RVs.

We performed a Monte Carlo analysis with the open-source Bayesian inference tool \textsc{MultiNest} v3.10 (e.g. \citealt{feroz13}), through the \textsc{pyMultiNest} \textsc{python} wrapper \citep{buchner14}, including the publicly available GP \textsc{python} module \textsc{GEORGE} v0.2.1 \citep{2015ITPAM..38..252A}. Our set-up included 500 live points and a sampling efficiency of 0.3. The use of a nested sampler allows for a robust Bayesian model comparison through the calculation of  marginal likelihood (or evidence) $\mathcal{Z}$ for each model with good accuracy, which is a crucial point for our analysis. In fact, we then compare the Bayesian evidence of a model containing only the correlated stellar activity term (that includes 5 free (hyper-)parameters), with that of a model including a planetary signal (that includes 9 or 11 free (hyper-)parameters, for a circular and eccentric orbit respectively) for a robust statistical analysis of our dataset.  

Since our goal is an independent confirmation of the presence in our data of the planetary signal detected by DO17, we adopted uninformative priors for all the free parameters except for $\theta$ to guarantee an unbiased analysis. The stellar rotation period can be reliably constrained using a Gaussian prior based on the result of a GP quasi-periodic regression of the H-$\alpha$ activity diagnostic time series ($\theta=2.7417\pm0.0007$~d), which is extracted from the same spectra used to derive the RVs. However, we adopted a more conservative value for the $\sigma$ of the prior, i.e. one order of magnitude larger that the uncertainty associated with the rotation period derived from the H-$\alpha$ time series. The orbital period of the planet was uniformly sampled up to 10 days.   

The results of the analysis for each of the different RV datasets are summarised in Table~\ref{tab:percentilesgp}. As an example of posterior distributions of model parameters, we show the corner plot for the GP+1 planet model (TERRA RVs) in Fig.~\ref{Fig:corneplotterragp1p}. The planetary signal detected by DO17 is not recovered in any of our datasets, and the marginal likelihoods always favor the 0-planet model. The GP regression is able to model effectively the stellar activity signal, as it can be seen by comparing the RMS of the original data to the RMS of the residuals. However, the latter are still above 100 \ms, which is nearly three times larger than the RMS of the residuals of DO17 (35 $\ms$). We do not have an explanation for this observed difference, that may be partly due to a higher level of activity of the star during our follow-up, or could be partly explained with the different wavelength ranges covered by HARPS and ESPaDOnS/Narval, with the latter reaching the NIR region where the RVs are expected to be less contaminated by stellar activity. 

We performed one more test by taking the RVs extracted with TERRA using \'echelle orders starting from no. 45, with the central wavelength  $\lambda=5463$\,\AA, i.e. we used a narrower region corresponding to a redder part of the spectra. The RVs have median uncertainties $\sigma_{\rm RV}$=10.2 \ms and RMS of 828 \ms, which are slightly less than that of the default dataset, in agreement with the evidence from STELLA data that the photometric variability is lower in \textit{I}-band than in \textit{V}-band (Fig. \ref{Fig:stellaterrafold}). Despite the lower scatter due to a reduced contribution from stellar activity, we did not find evidence for the planetary sign.

Leaving the eccentricity unconstrained does not provide any improvement to the fit (for instance, we get $e_{\rm b}=0.46^{+0.36}_{-0.31}$ and $\ln\mathcal{Z}=-992.3$ for the TERRA dataset). We also used a looser prior for $P_{\rm b}$, increasing the upper limit to 100 d, in order to explore the possible presence of longer period planets. Even so, we did not find evidence for any significant signal in the data.

\begin{sidewaystable*}
       \caption{Results of the GP regression analysis applied to RVs extracted with different pipelines. Models with and without zero-eccentricity Keplerians were considered. MAP values for the Keplerian parameters $K_{\rm b}$ and $P_{\rm b}$ are given in parenthesis.}
       \label{tab:percentilesgp}
       \small
 \begin{tabular}{llcccccc}
           \hline
           \hline
            \noalign{\smallskip}
             & & \multicolumn{2}{c}{TERRA} & \multicolumn{2}{c}{DRS} & \multicolumn{2}{c}{R20} \\
            \noalign{\smallskip}
            \hline
            \noalign{\smallskip}
             & & $N=0$ planets &  $N=1$  & $N=0$  & $N=1$  & $N=0$  & $N=1$  \\
            \hline 
            \noalign{\smallskip}
             Parameter &  Prior & Best-fit value  & Best-fit value & Best-fit value & Best-fit value & Best-fit value & Best-fit value \\
            \noalign{\smallskip}
            \hline
            \noalign{\smallskip}
            \hline
            \noalign{\smallskip}
            \textbf{GP hyper-parameters} & & & & & & &\\
            \noalign{\smallskip}
            \hline
            \noalign{\smallskip}
            $h$ [m$\,s^{-1}$] & $\mathcal{U}$(0,1500) & 867$_{\rm -128}^{\rm +179}$ & 869$_{\rm -105}^{\rm +151}$ & 1167$_{\rm -159}^{\rm +173}$ & 1176$_{\rm -132}^{\rm +154}$ & 955$_{\rm -133}^{\rm +182}$ & 955$_{\rm -111}^{\rm +153}$ \\
            \noalign{\smallskip}
            $\lambda$ [days] & $\mathcal{U}$(0,1000) & 229$^{\rm +29}_{\rm -26}$ & 228$_{\rm -22}^{\rm +26}$ & 241$_{\rm -31}^{\rm +41}$ & 242$_{\rm -28}^{\rm +36}$ & 229$^{\rm +36}_{\rm -31}$ & 225$^{\rm +29}_{\rm -26}$ \\
            \noalign{\smallskip}
            $w$ & $\mathcal{U}$(0,1) & 0.37$\pm$0.04 & 0.37$^{\rm +0.04}_{\rm -0.03}$ & 0.29$\pm$0.03 & 0.29$\pm$0.02 & 0.34$^{\rm +0.04}_{\rm -0.03}$ & 0.34$\pm$0.03 \\ 
            \noalign{\smallskip}
            $\theta$ [days] & $\mathcal{N}$(2.742,$\sigma=0.007$) & 2.7409$\pm$0.0004 & 2.7409$\pm$0.0004 & 2.7411$\pm$0.0004 & 2.7411$\pm$0.0004 & 2.7410$^{\rm +0.0005}_{\rm -0.0004}$ & 2.7410$\pm$0.0004 \\
            \noalign{\smallskip}
            \hline
            \noalign{\smallskip}
            $\sigma_{\rm jit,\: HARPS-N}$ [m$\,s^{-1}$] & $\mathcal{U}$(0,500) & 117$_{\rm -10}^{\rm +12}$ & 115$_{\rm -9}^{\rm +10}$ & 194$_{\rm -17}^{\rm +19}$ & 192$_{\rm -15}^{\rm +17}$ & 108$_{\rm -13}^{\rm +14}$ &  106$_{\rm -11}^{\rm +14}$ \\ 
            \noalign{\smallskip}
            $\gamma_{\rm HARPS-N}$ [$\ms$] & $\mathcal{U}$(-1000,+1000) [TERRA; R20] & & -207$_{\rm -259}^{\rm +281}$ & & 17329$^{+320}_{-311}$ & & -88$_{\rm -283}^{\rm +282}$ \\
            & $\mathcal{U}$(16500,18500) [DRS] \\
            \noalign{\smallskip}
            \hline
            \noalign{\smallskip}
            \textbf{Planet parameters} & & & & & & & \\
            \noalign{\smallskip}
            \hline
            \noalign{\smallskip}
            $K_{\rm b}$ [m$\,s^{-1}$] & $\mathcal{U}$(0,100) & & 25.4$_{\rm -17.0}^{\rm +21.0}$ (48.1) & & 37$_{\rm -24}^{\rm +30}$ (77) & & 25.7$_{\rm -16.9}^{\rm +22.7}$ (75.8) \\
            \noalign{\smallskip}
            $P_{\rm b}$ [days] & $\mathcal{U}$(0,10) & & 4.0$_{\rm -1.5}^{\rm +4.5}$ (1.4) & & 5.9$_{\rm -3.5}^{\rm +2.3}$ (3.3) & & 3.9$_{\rm -1.8}^{\rm +3.9}$ (3.6) \\ 
            \noalign{\smallskip}
            $T_{\rm b,\:conj}$ [BJD$-2\,450\,000$] & $\mathcal{U}$(8840,8855) & & 8847.0$_{\rm -4.2}^{\rm +5.2}$ & & 8846.9$_{\rm -4.3}^{\rm +4.9}$ & & 8847.1$_{\rm -4.4}^{\rm +5.1}$ \\
            \noalign{\smallskip}
            \noalign{\smallskip}
            \hline
            \noalign{\smallskip}
            \hline
            \noalign{\smallskip}
            Bayesian evidence $\ln \mathcal{Z}$ & & $-990.6$ & $-992.0$ & $-1058.3$ & $-1059.3$ & $-1002.8$ & $-1004.3$  \\ 
            \noalign{\smallskip}
            RMS of the residuals [\ms] &  & 105 & 102 & 157 & 161 & 109 & 126 \\
            \noalign{\smallskip}
            \hline
            \hline
 \end{tabular}
\end{sidewaystable*}

\subsection{Gaussian process RV modelling jointly with ancillary activity indicators}
\label{subsec:vineshgpmodeling}

To investigate in more detail the interplay between RV variations and stellar activity, we applied a more sophisticated GP-based approach. We analysed all the different RV datasets in Table~\ref{table:rvproperties} using the framework described by \cite{vinesh15}. The RVs were fitted jointly with the DRS CCF asymmetry indicators BIS and FWHM. In short, this GP framework assumes that all observed stellar activity signals are generated by some underlying latent function $G(t)$ and its derivatives; this function, which is not observed directly, is modelled with a Gaussian process with a quasi-periodic covariance function. $G(t)$ and its derivative are then allowed to manifest (using physically-motivated relationships) in all observable, activity-sensitive time series, while Keplerian terms for one or more planets are incorporated into the RVs only. This GP-based approach to model RVs jointly with activity indicators could enable us more reliable planet characterisation compared to traditional approaches that assume simple parametric forms for the stellar signals, or that try to exploit simple correlations between RVs and activity indicators. 

In an effort to detect \object{V830 Tau~b} in our RVs, we combined the GP-based activity framework with models including either one or no planet(s). We placed uninformative priors on all standard parameters in \cite{vinesh15} framework, as well as on the planet parameters, except for the period, which we constrained to $4.93\pm0.25$~d (i.e., looking now for an expected signal at that period, 
rather then doing a blind search over all the periods). To compute model posteriors and Bayesian evidences, we employed \textsc{PolyChord} \citep{handley15}, which is a state-of-the-art nested sampling algorithm, and an efficient alternative to \textsc{MultiNest} designed to work especially with very high dimensional parameter spaces. 

In brief, we found that the Bayes factor for the 1-planet models vs.\ the 0-planet models ranged from 1.26 to 5.52, depending on the RV extraction algorithm used (e.g. DRS vs.\ TERRA): in no instance, then, was a 1-planet model strongly favoured. More decisively, the RV semi-amplitude for the $4.93$~d-period Keplerian was in all cases consistent with zero within $1\sigma$, indicating the non-detection of \object{V830 Tau~b}.

In various other tests where we used the same GP framework but replaced the narrow planet period prior with an uninformative one, 1-planet models were always rejected outright compared to the 0-planet case. The period posteriors had little probability density around $\sim4.9$~d, and the semi-amplitudes associated with $\sim4.9$~d-period samples were clustered tightly around zero. These results, even more strongly than in the case of the narrow period prior, indicated a non-detection of \object{V830 Tau~b}.
\subsection{Modelling the RV variations from the observed photometric modulation}
\label{subsec:ffprime}
Wide-band photometry can be used to map the longitudinal distribution of active regions on the surface of an active star and to predict, to some extent, the activity-induced RV variations \citep[e.g.,][]{lanza11}. 

\begin{figure*}
 \centering{
 \includegraphics[scale=0.5,angle=270]{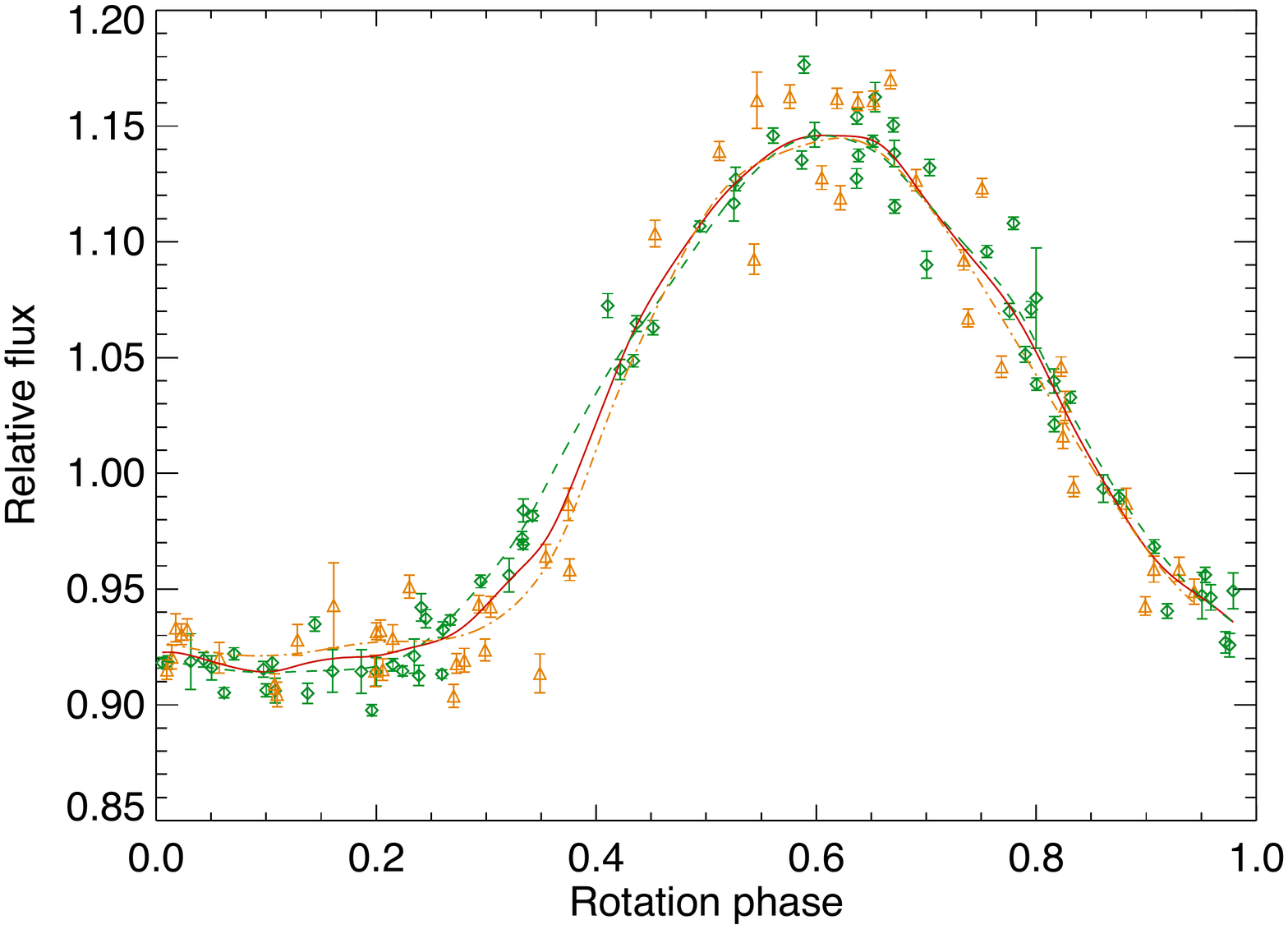}} 
 \vspace*{-10mm}
   \caption{The V-band optical lightcurves of \object{V830 Tau} with the flux plotted vs.~the rotation phase. The data points collected between BJD 58393.742 and 58607.363 are plotted as green open diamonds and their kernel regression {vs.} the rotational phase is given by the dashed green line, while the data points collected between BJD 58765.645 and 58916.387 are plotted as open orange triangles and their regression is given by the dot-dashed orange line. The solid red line is the regression to the whole \textit{V}-band photometric dataset vs.~the rotation phase. }
              \label{V-band_light_curves_pub}%
\end{figure*}

In Fig.~\ref{V-band_light_curves_pub} we plot the two seasons of V-band optical photometry of V830~Tau vs.~the rotation phase together with continuous interpolations obtained for the individual seasons as well as for the dataset as a whole. The rotation phase is computed assuming a constant rotation period of $P_{\rm rot} = 2.7409$~d. To compute the interpolations, we performed a Kernel Regression (KR), that is, a locally linear regression of the RV vs.\ the phase giving decreasing weights to the data points that are more distant in phase from the given data point for which the regression value is to be computed (see Sect.~\ref{subsec:kernreg} for details).  We see that only minor changes occur between the two seasonal light curves, although they are separated by more than $300$~d. This indicates  that the photospheric brightness inhomogeneities are very stable in \object{V830 Tau}. Therefore, we consider our \textit{V}-band photometric dataset as a whole, thus obtaining a more continuous phase coverage for our subsequent analysis. 
An analogous approach can be applied to the \textit{I}-band light curves, but we focus on the \textit{V}-band light curves because they show a flux modulation of greater amplitude, thanks to the higher starspot contrast in the optical, which permits a more precise RV reconstruction. 

To compute the activity-induced RV variations, we apply the so-called FF$^{\prime}$ method introduced by \cite{aigrain12}. It accounts for both the variation $\Delta RV_{\rm rot}$ induced by the spectral line distortions produced by surface brightness inhomogeneities, the visibility of which is modulated by stellar rotation,  and the variation $\Delta RV_{\rm c}$ produced by the inhibition of surface convection in the regions where photospheric magnetic fields are more intense, that reduces the local convective blueshifts of spectral lines. We adopt the following expressions for the two components:
\begin{equation}
\Delta RV_{\rm rot} (\phi) =  \frac{A}{F_{0}} \frac{dF(\phi)}{d\phi} \left[\frac{F(\phi)}{F_{0}} -1 \right] 
\end{equation} 
and
\begin{equation}
\Delta RV_{\rm c} (\phi) = B \left[\frac{F(\phi)}{F_{0}}-1\right]^{2},
\end{equation}
where $\phi$ is the rotation phase, $F(\phi)$ the interpolated \textit{V}-band flux at phase $\phi$, $F_{0}$ the flux in the absence of spots (that we take equal to the maximum flux along the interpolated light curve), and $A$ and $B$ two coefficients to be determined together with the RV offset $RV_{0}$ between the model and the observations by minimising the $\chi^{2}$. This is computed as the sum of the squares of the residuals between the model radial velocities and the observations, normalised by the respective standard deviations of the RV measurements. 

\begin{figure*}
 \centering{
 \includegraphics[scale=0.5,angle=270]{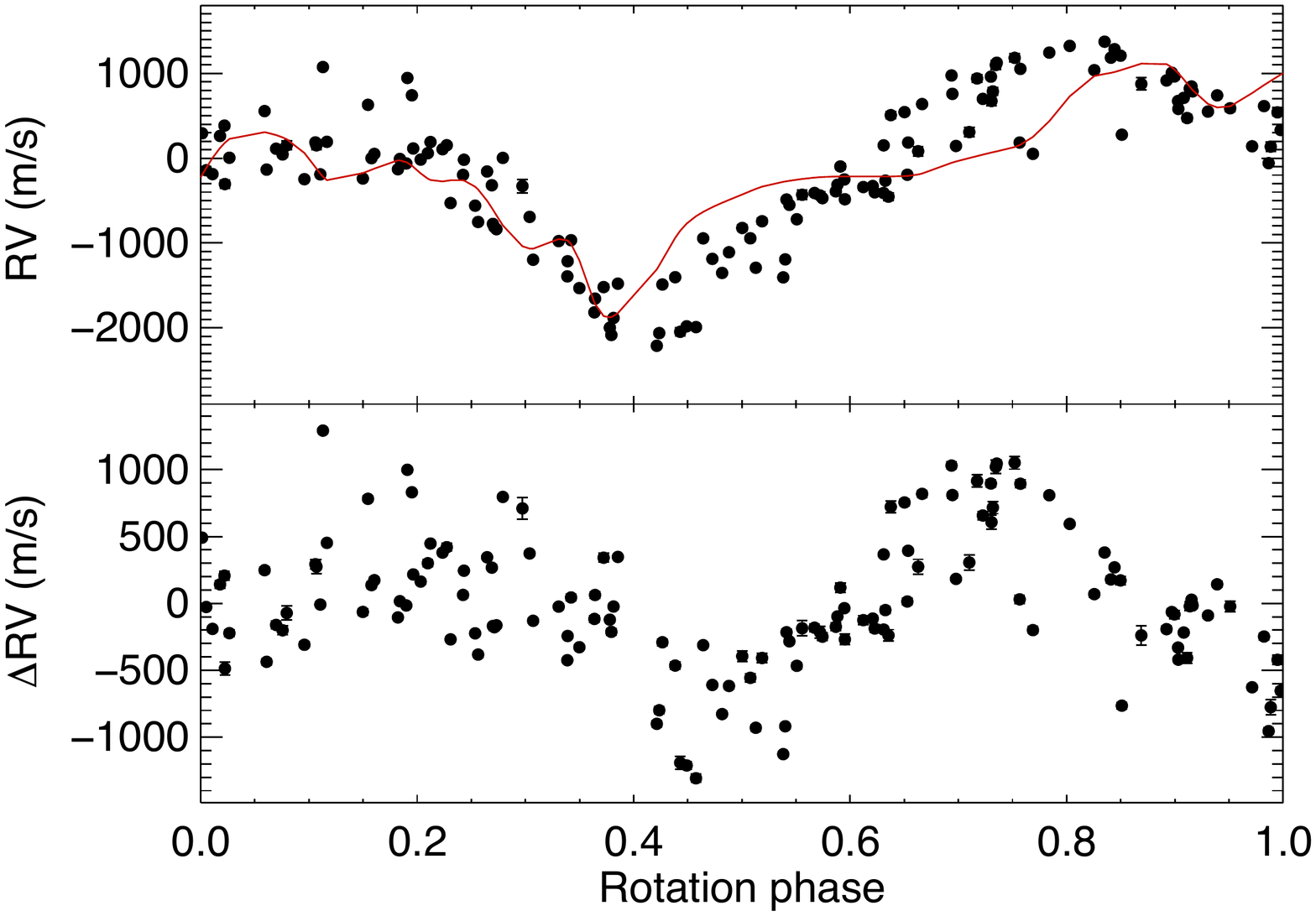}} 
 \vspace*{-10mm}
   \caption{Top panel: RVs obtained with the TERRA procedure vs.~time (filled dots). The model RV variation as derived from the red solid interpolation in Fig.~\ref{V-band_light_curves_pub} by means of the FF$^{\prime}$ method is superposed (red solid line). Bottom panel: residuals between the observed TERRA RV and the RV predicted by the FF$^{\prime}$ model vs.~time. In both panels the RV error bars have a size comparable with that of the data points in most of the cases.}
              \label{RV_obs_model_comparison_TERRA_pub}%
\end{figure*}

The $\chi^{2}$ minimisation with respect to the RV extracted with the TERRA procedure gives the model plotted as a red solid line in the top panel of Fig.~\ref{RV_obs_model_comparison_TERRA_pub}, the residuals of which are plotted in the bottom panel of the same figure and have a standard deviation of 509.33~m~s$^{-1}$, while the original RV time series in the top panel has a standard deviation of 874.99~m~s$^{-1}$. The model RV variations are dominated by the effect of the brightness inhomogeneities with a mean value of $|\Delta RV_{\rm rot}| $ equal to 4.93 times the mean value of $|\Delta RV_{\rm c}|$ as one could expect given the rather large rotational broadening of the spectral lines ($v \sin i \sim 30$~km~s$^{-1}$) and the large area occupied by starspots on V830~Tau. Similar results are obtained with the RV extracted by the HARPS-N pipeline DRS, although with a larger RV dispersion and greater residuals for the FF$^{\prime}$ model. 

In conclusion, our RV model based on wide-band photometry does not adequately reproduce the observed RV variations of V830~Tau, probably because the pattern of surface brightness inhomogeneities is much more complex than the simple spot distribution assumed by the FF$^{\prime}$ model. This is clearly indicated by the Doppler imaging maps of DO17, who warned about the limitations of any reconstruction of the RV variations based solely on the photometry for this very active star. 

\subsection{Kernel regression analysis of the RV time series}
\label{subsec:kernreg}

In another, complementary analysis of the V830~Tau RVs, we first tried to remove the rotational modulation produced by stellar activity. The evolution timescale of the surface features produced by magnetic activity is comparable with the time span of the observations in individual seasons, thus an effective method to reduce the activity-induced RV modulation is to put the RV data in phase for each season and make a regression vs.~the rotation phase. By subtracting such a regression, we remarkably reduce the activity-induced RV variations. We consider our three seasons of RV data, the first between BJD 58044.6623 and 58192.3573, the second between BJD 58341.7344 and 58566.3847, and the third between BJD 58804.4827 and 58924.4063, and put each of them in phase assuming a constant rotation period $P_{\rm rot} = 2.7409$~d. 

To compute the regression of the RV vs.~the rotation phase, we performed a KR. More precisely, to compute the regression value for the $i$-th data point  observed at time $t_{i}$,  corresponding to rotation phase $\phi_{i}$, we performed a linear regression of the RV versus the phase over all the data points giving them a weight 
\begin{equation}
w_{j} \propto \exp \left\{ - \left[ \left( \frac{\phi_{i}-\phi_{j}}{h_{\rm \phi}} \right)^{2} + \left( \frac{t_{i} -t_{j}}{h_{\rm t}} \right)^{2} \right] \right\}, 
\end{equation}
where $\phi_{j}$ is the rotation phase of the generic $j$-th data point observed at time $t_{j}$, while  $h_{\rm \phi}$ and $h_{\rm t}$ are the so-called bandwidths; they  govern the decrease of the weight $w_{j}$ of the generic $j$-th data point as it becomes more and more distant from the considered $i$-th data point. More details on the KR implementation can be found in  \citet{lanza18}, \citet{lanza19}, and references therein.  

\begin{figure*}
 \centering{
 \includegraphics[scale=0.5,angle=270]{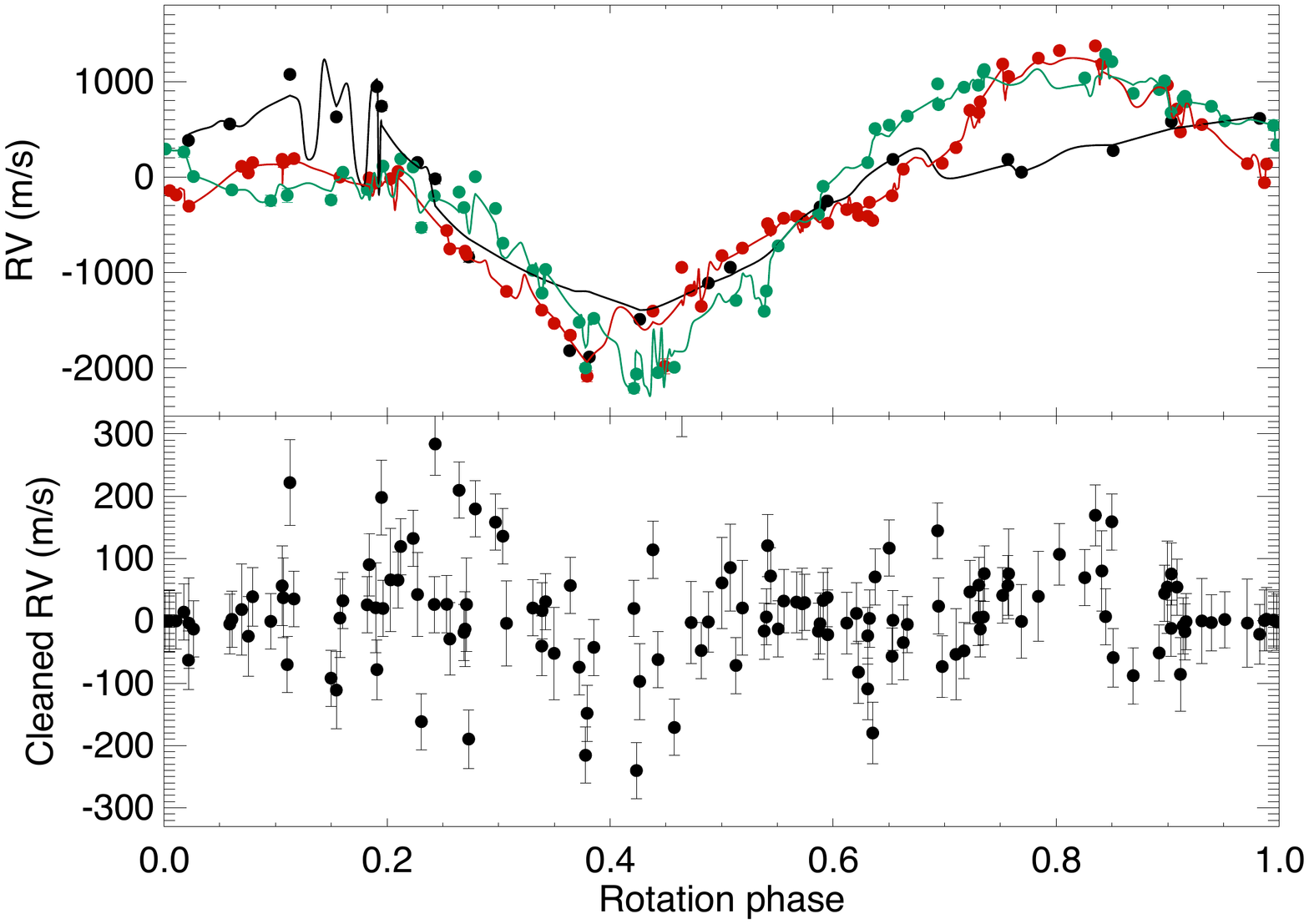}} 
 \vspace*{-10mm}
   \caption{Top panel: The RV time series of \object{V830 Tau} as extracted with the TERRA procedure vs.~the rotation phase (filled dots). Different  colours indicate data collected in different seasons: black dots indicate data points collected between BJD 58044.6623 and 58192.3573 (first season),  red dots between BJD 58341.7344 and 58566.3847 (second season), while  green dots between 58804.4827 and 58924.4063 (third season). The KR performed vs.~the rotation phase and time in the first season is indicated by the solid black line, in the second season by the solid red line, and in the third season by the green solid line. The RV errorbars are smaller than the size of the plotted dots. Bottom panel: RV residuals obtained by subtracting the seasonal KRs  from the corresponding data points. }
              \label{GAPS_phased_timeseries}%
\end{figure*}

The results of the application of KR to our seasonal RV datasets extracted with the TERRA procedure are illustrated in Fig.~\ref{GAPS_phased_timeseries}. The whole TERRA RV time series consists of 144 data points with a standard deviation of 874.99 m~s$^{-1}$ and is plotted in the top panel with different colours indicating data collected in different seasons. The same colour code is used to plot the corresponding  seasonal KRs. The residual time series obtained by subtracting the seasonal KRs  has a standard deviation of 123.91 m~s$^{-1}$ and is plotted in the bottom panel.  We shall refer to this residual RV time series as the cleaned RV time series. The mean bandwidths over the three seasons are $h_{\rm \phi} = 0.095$ and $h_{\rm t} = 19.73$~d. 

To further reduce the RV scatter, we considered the indicators of the shape of the CCF and the chromospheric index $\log R^{\prime}_{\rm HK}$ that measures the excess flux in the core of the Ca II H\&K lines produced by the non-radiative heating controlled by magnetic activity. In addition to the commonly used BIS index, our suite of CCF shape indicators included the contrast of the CCF, its full width at half maximum, $\Delta V$, and  $V_{\rm asy (mod)}$ introduced in \citet{lanza18}. We performed KRs of the cleaned RV time series vs.~each of these indicators and the time. Additionally, we performed a further KR with respect to the rotational phase and time. In all the cases, as in \citet{lanza18}, a 3-$\sigma$ clipping was applied by performing a preliminary KR to exclude possible outliers. 

\begin{figure*}
 \centering{
 \includegraphics[scale=0.5,angle=270]{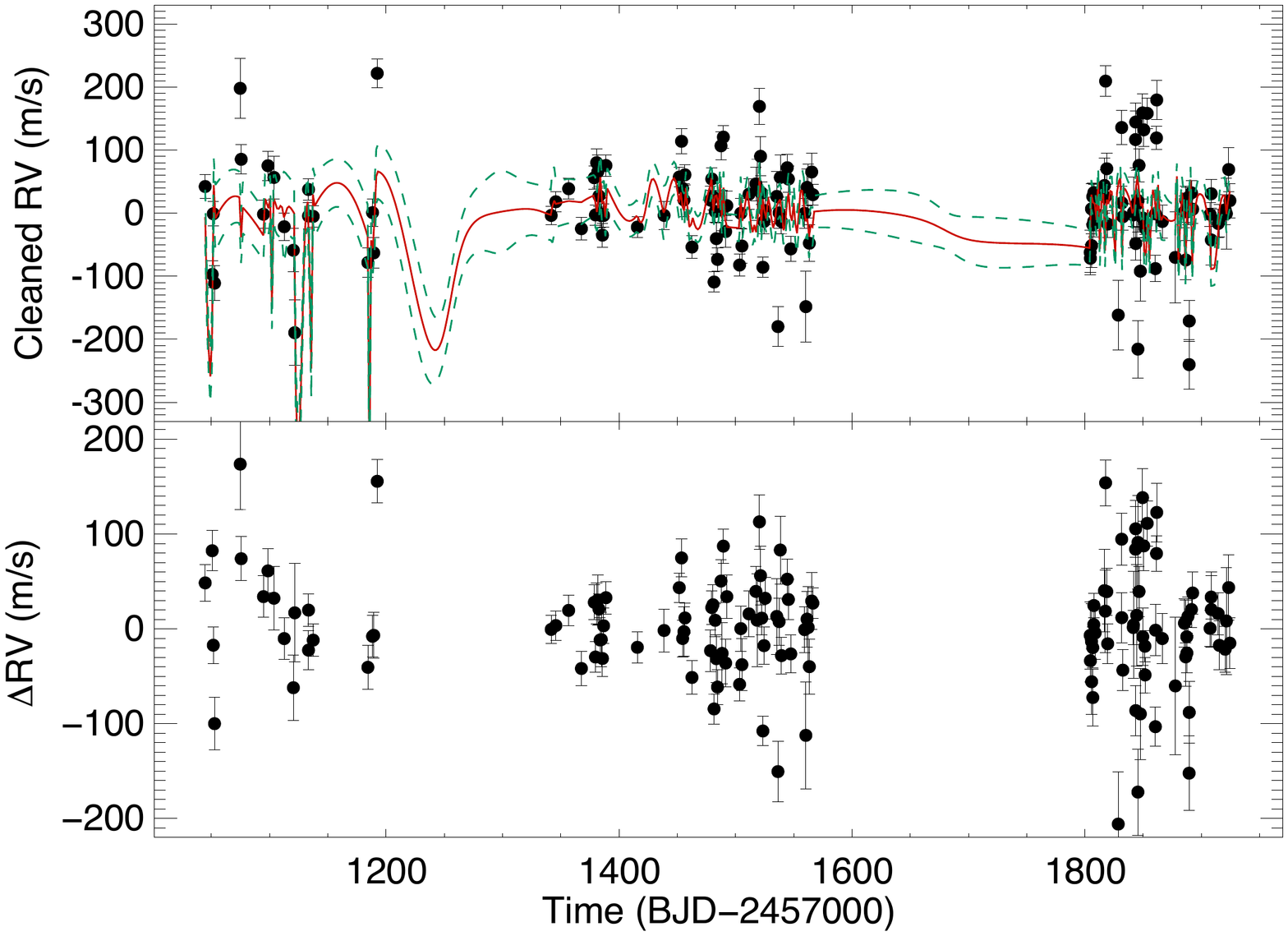}} 
 \vspace*{-10mm}
   \caption{Top panel: The cleaned RV time series of \object{V830 Tau} vs.~time as obtained after the first KR with respect to the rotation phase and time (filled dots). The red solid line indicates a further KR with respect to the phase and time that gives the largest reduction of the standard deviation of the residuals in comparison with the KRs computed with respect to the CCF indicators or the chromospheric index $\log R^{\prime}_{\rm HK}$. The dashed green lines indicate the $1\sigma$ confidence range of the KR \citep[see][for details]{lanza18}. Bottom panel: residuals of the KR in the top panel vs.~time, with a standard deviation of 65.13~m~s$^{-1}$.  }
              \label{GAPS_timeseries_cleaned}%
\end{figure*}

None of these KRs gave a significant reduction of the standard deviation of the data points as measured by the Fisher-Snedecor $F$ statistics \citep[see][]{lanza19}; therefore we simply consider the one giving the smallest standard deviation of the residuals. That turned out to be the KR with respect to stellar rotation phase and time, probably because the CCF indicators lose most of their power when the CCF is strongly distorted as in the case of a very active star such as V830~Tau, while the chromospheric index $\log R^{\prime}_{\rm HK}$ is not strongly correlated with the photospheric activity mainly responsible for the RV variations. This is supported by the analysis of H$\alpha$ emission performed by DO17. Note that the second KR with respect to phase and time is different from the first KR applied to obtain the cleaned RV time series because it has a longer time bandwidth $h_{\rm t} = 68.15$~d, although the phase bandwidth is the same $h_{\rm \phi} = 0.095$. The standard deviation of the residuals after this second KR is 65.33 m~s$^{-1}$ for a total of 140 data points because the 3-$\sigma$ clipping excluded four outliers. The KR applied to the cleaned time series and the obtained residuals are shown in Fig.~\ref{GAPS_timeseries_cleaned}. 

\begin{figure}
\hspace*{-7mm}
 \centering{
 \includegraphics[width=0.8\hsize,angle=270]{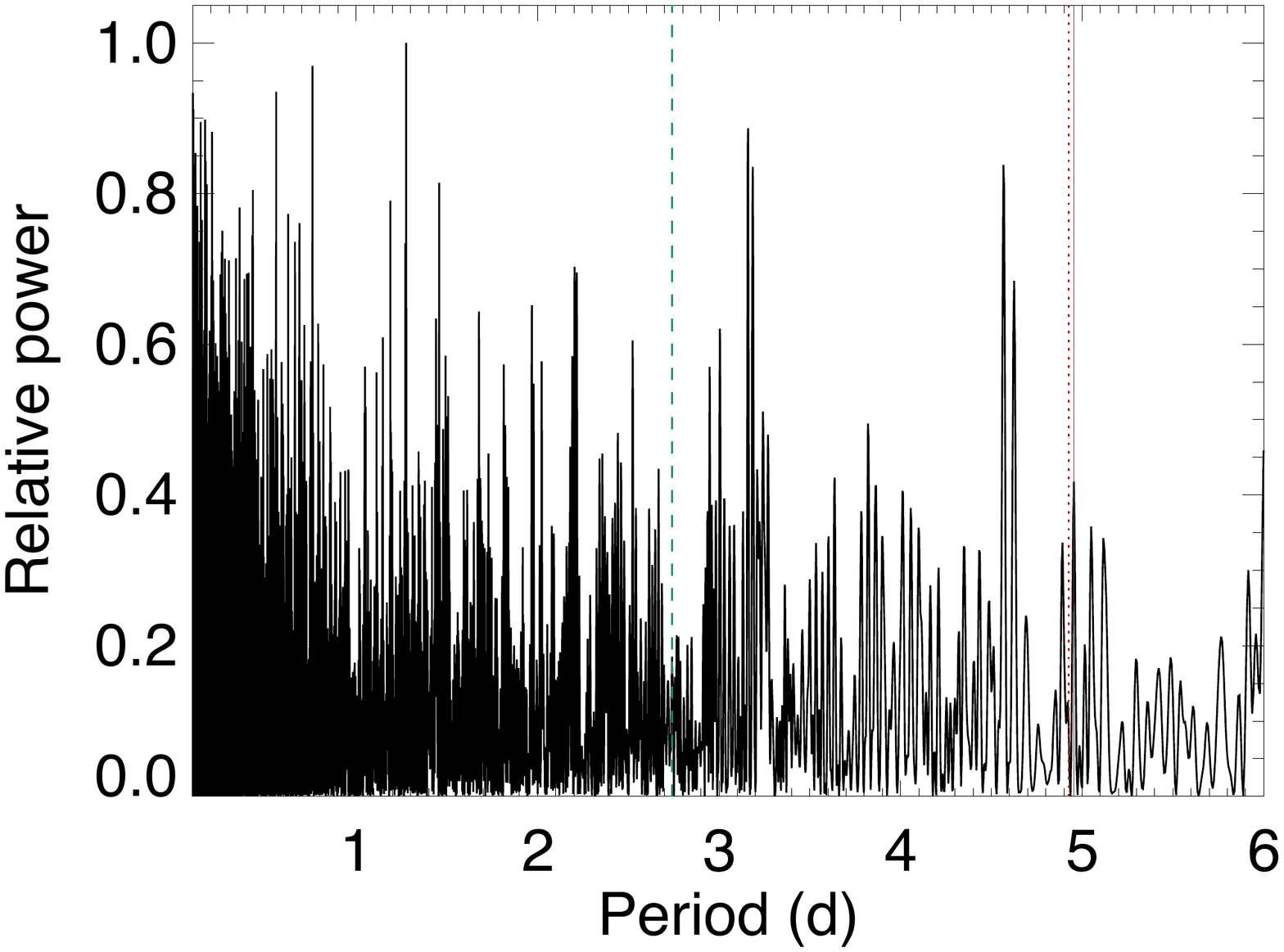}} 
 \vspace*{-10mm}
   \caption{Generalized Lomb-Scargle periodogram of the residual RV time series of \object{V830 Tau} in the bottom panel of Fig.~\ref{GAPS_timeseries_cleaned}. The power is normalised to its maximum in the considered period interval and is plotted vs.~the period itself. The vertical green dashed line marks the rotation period, while the red dotted line indicates the orbital period of the planet proposed by \citet{donati17}.  }
              \label{periodogram_GAPS_extended}%
\end{figure}

In Fig.~\ref{periodogram_GAPS_extended}, we plot a GLS periodogram of the residual RV time series in the bottom panel of Fig.~\ref{GAPS_timeseries_cleaned}. The false-alarm probability corresponding to the highest peak is 0.642 as given by the ana\-ly\-ti\-cal formula of \citet{zech09}; therefore, there is no indication of significant periodicities in the explored period range. The peak closest to the orbital period of DO17 falls at 4.9545~d. By fitting a sinusoid with this period to the RV time series, we find a semi-amplitude of only 18.63~m~s$^{-1}$,  much smaller than the orbital RV semi-amplitude of $68 \pm 11$~m~s$^{-1}$ reported by DO17. 

The possibility that our two successive KRs with respect to stellar rotation phase and time might have removed a signal at the period of the putative planet appears to be very low because both their time bandwidths $h_{\rm t}$'s are significantly longer than the period of 4.927~d. We acknowledge that the approach we used for our KR analysis is not completely appropriate from a statistical point of view because the activity and the sinusoidal fit should be performed simultaneously rather than applying the GLS to the KR residuals (see, e.g., \citealt{anglada15}). Nevertheless, it is much simpler and can be adequate for an exploratory analysis such as that presented here. 

Therefore, we conclude that even with the alternative and complementary KR technique we cannot detect a significant signal at the period of V830\,Tau\,b.


\section{Planet detection through injection-retrieval simulations}
\label{sec:simulations}

The lack of the exoplanetary signal reported by DO17 in our time series needs to be investigate in terms of effective detectability in presence of such a high level of stellar activity.
To this purpose, we devised GP-based simulations to test the feasibility of retrieving the planetary signal claimed by DO17, once this is injected into our data (adopting the TERRA dataset), following two different approaches. 

\subsection{Detection sensitivity by direct injection of the planetary signal into the data}
\label{sect:simu_inj_real_data}
\begin{figure}
  \centering
   \includegraphics[width=0.8\hsize]{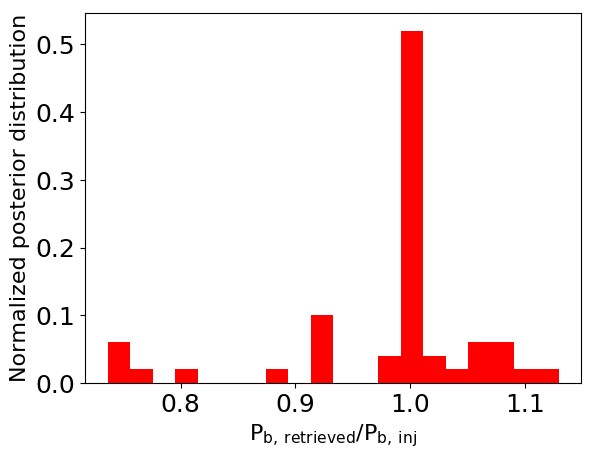}\\
   \includegraphics[width=0.8\hsize]{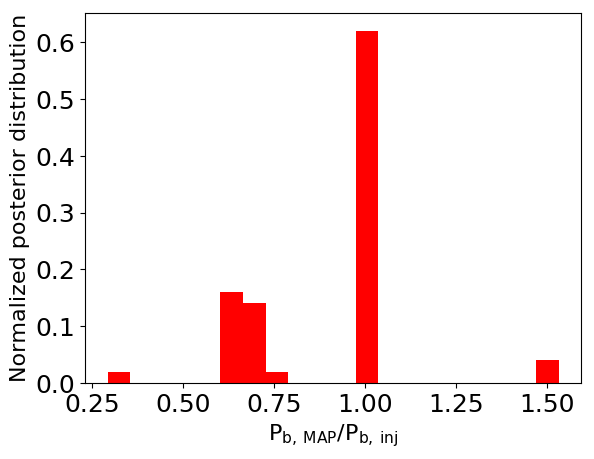}
    \caption{Results for the simulations with direct injection of the planetary signal in the real data. }
    \label{Fig:simu_1_dist}
\end{figure}

The first set of simulations was built by direct injection of the planetary signal in our data, after randomly drawing the parameters from normal distributions determined from the DO17 results ($K_{\rm b}=68\pm11$~\ms, $P_{\rm b}=4.927\pm0.008$~d, $T_{\ conj,\:b}=2457360.523\pm0.124$~BJD). We produced 50 mock datasets, which were analysed with a GP regression including a sinusoid to fit the planetary signal, and with the same set-up used for fitting the real data (e.g., adopting a uniform prior $\mathcal{U}$(0,10) days for $P_{\rm b}$). Then, as a figure of merit, we inspected the distribution of the $P_{\rm b,\,retrieved}$/$P_{\rm b,\:inj}$ ratio between the 50$^{th}$-percentile ($P_{\rm b,\,retrieved}$) of each $P_{\rm b}$ posterior and the corresponding injected orbital period $P_{\rm b,\,injected}$. We derived a similar distribution using the maximum a posteriori probability (MAP) values $P_{\rm b,\,MAP}$ in place of $P_{\rm b,\,retrieved}$. Both histograms are shown in Fig.~\ref{Fig:simu_1_dist}. For more than $50\%$ of the simulated datasets we retrieved an accurate value for $P_{\rm b}$ within the range 4.9<$P_{\rm b}$<5.0 d, with a median significance level of 9.5$\sigma$ ($20\%$ of the whole dataset having a significance higher than 100$\sigma$). For all the datasets of this sub-sample $P_{\rm b,\,MAP}$ corresponds to $P_{\rm b,\:inj}$ with high accuracy, except one dataset for which $P_{\rm b,\,MAP}=3.24$~d. About this sub-sample, we were not able to recover the semi-amplitude of the planetary signal with the same degree of accuracy, getting 0.98 and 0.28 for the median and RMS of the $K_{\rm b,\,retrieved}$/$K_{\rm b,\:inj}$ ratio, respectively.
If we assume the MAP values as an estimate for the orbital period, the percentage of the cases for which we can claim an accurate recovered $P_{\rm b}$ increases to $\sim$60$\%$. Within this sub-sample, we found 5 datasets for which $P_{\rm b,\,retrieved}$ is not in the range 4.9-5.0 d, while the median $K_{\rm b,\,retrieved}$ is 20$\%$ overestimated.

We tested our ability to recover the planetary signal also by setting the semi-amplitude to a couple of illustrative larger values $K_{\rm b}=100$ and 130 $\ms$, and increasing the upper prior bound to 200 $\ms$. For both cases, this time we considered only one realisation. For the first case, we retrieved $K_{\rm b}=89^{+18}_{-36}$~$\ms$ (MAP value $K_{\rm b}=103$~$\ms$) and $P_{\rm b}=4.915\pm0.004$~d ($P_{\rm b,\:injected}=4.921$~d); the model including the planetary signal was only moderately favoured over the model with just the correlated stellar activity signal, with a Bayes factor of about 5, which is not enough to claim a statistically significant detection. When using $K_{\rm b}=130$~$\ms$, the planetary signal was much better recovered ($K_{\rm b}=136^{+14}_{-16}$~$\ms$ and $P_{\rm b}=4.932\pm0.002$~d, with $P_{\rm b,\:injected}=4.933$~d), and with a high significance (Bayes factor of about 9$\times10^{6}$). This simple test demonstrates that we can reliably detect the planet when the semi-amplitude of the injected signal is greater than the RMS of the RV residuals of the real data, after removing the quasi-periodic activity signal (see Table~\ref{tab:percentilesgp}).    

\subsection{Injection-recovery simulations under a more general scheme}
\label{sect:simu_inj_varying_hyper}
For the second set of simulations, we kept the same time stamps of the real data, and generated 100 mock datasets as follows. We added the planetary signal of DO17 to the best-fit stellar activity signal that we determined through a GP regression including one planet. We used the error bars of each GP hyper-parameter and of the uncorrelated jitter $\sigma_{\rm jit}$ to randomly draw arrays of parameters from a multi-dimensional normal distribution. The arrays of hyper-parameters were used to generate the quasi-periodic stellar activity term, to which we added the planetary signal (with $K_{\rm b}$, $P_{\rm b}$ and $T_{\rm conj,b}$ drawn from normal distributions, as done before). Finally, we added a randomly generated `white noise' term with RMS equal to that of the residuals of the real data (105 \ms), and the so obtained RV values were randomly shifted within the internal errors given as $\sqrt{\sigma^{2}_{\rm RV}+\sigma^{2}_{\rm jit}}$ (with $\sigma^{2}_{\rm jit}$=115\ms), still adopting a normal distribution. 

\begin{figure}
   \includegraphics[width=\hsize]{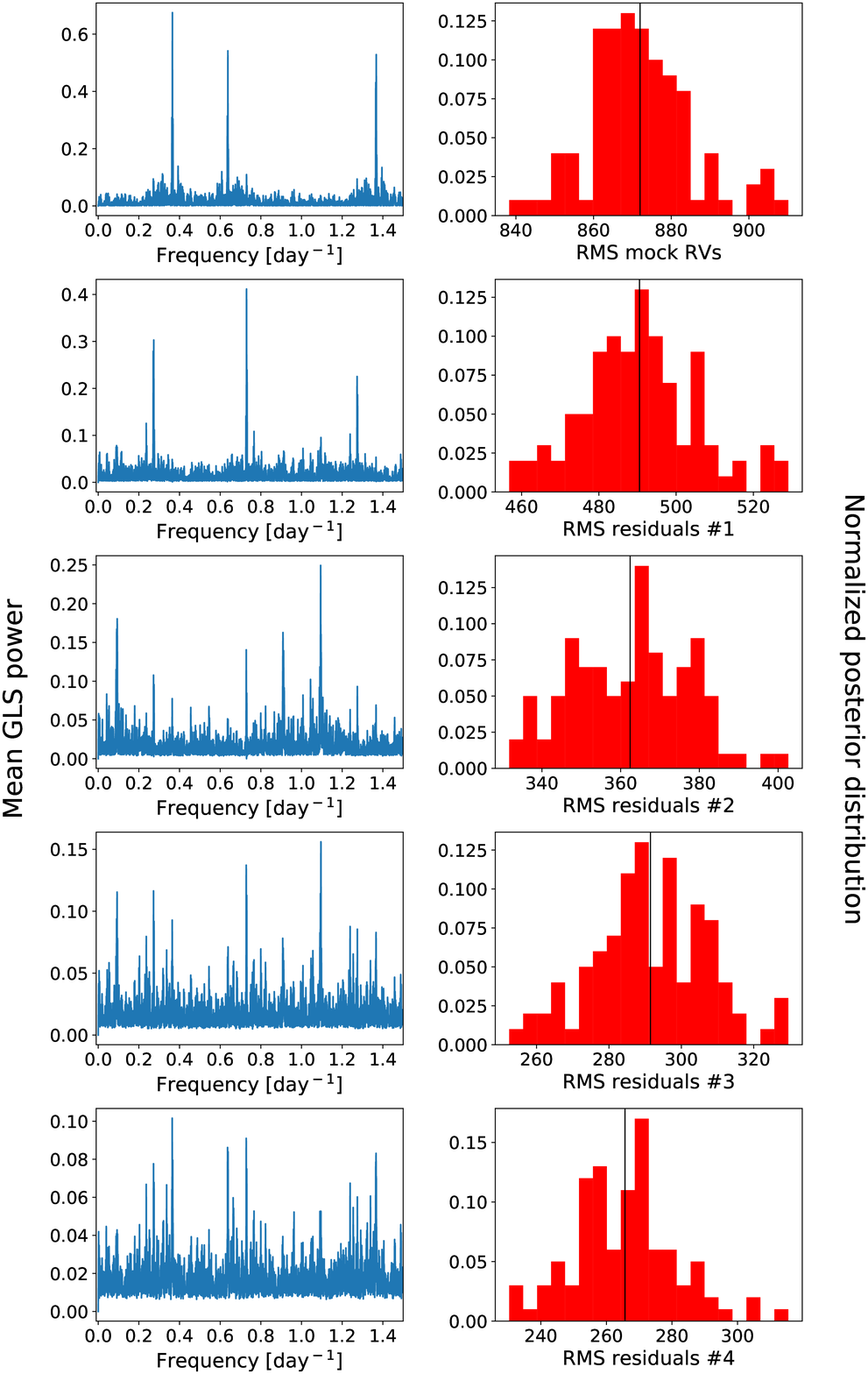}
    \vspace*{-10mm}
    \caption{Statistical properties of the simulated dataset described in Section \ref{sect:simu_inj_varying_hyper}. The average GLS periodograms and the distributions of the RMS of the data (original data and residuals) are shown in the left and right columns, respectively. Each row, starting form the second, refers to residuals determined through iterative pre-whitening.}
    \label{Fig:sim2properties}
\end{figure}

Statistical properties of the simulated datasets are shown in Fig.~\ref{Fig:sim2properties}. They should be compared with those of the real TERRA RVs from which they were derived (see periodograms and RMS values in Fig.~\ref{Fig:glsrecursiveterra}). It can be seen that the mean GLS periodograms (left column) and the distributions of the RMS of the simulated data (original data and residuals determined through iterative pre-whitening) are on average well consistent with those of the real data.

\begin{figure*}
  \begin{subfigure}{.5\textwidth}
   \centering
   \includegraphics[width=0.85\hsize]{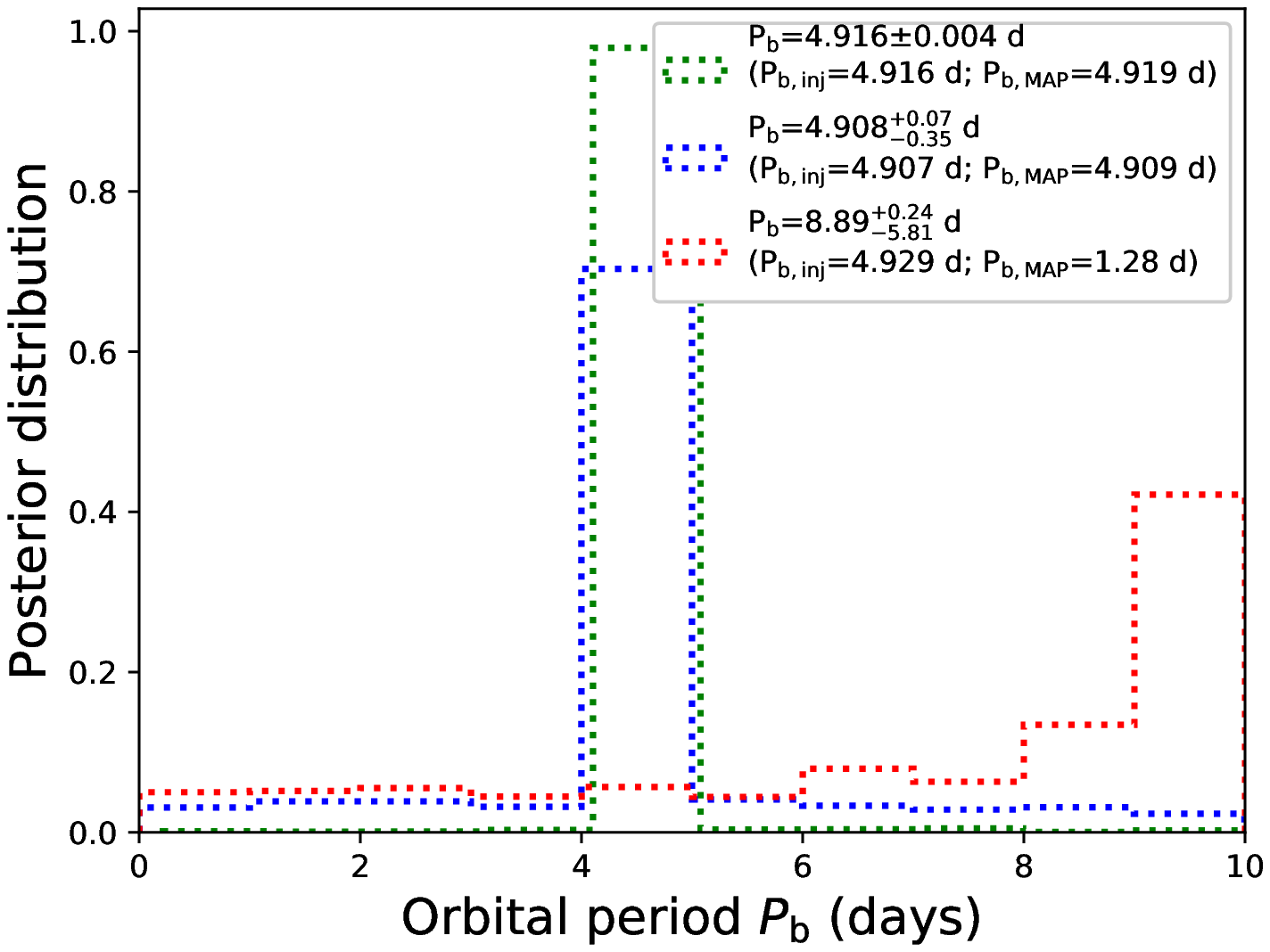}
   \caption{}
  \end{subfigure}%
  \begin{subfigure}{.5\textwidth}
  \centering
  \includegraphics[width=0.85\hsize]{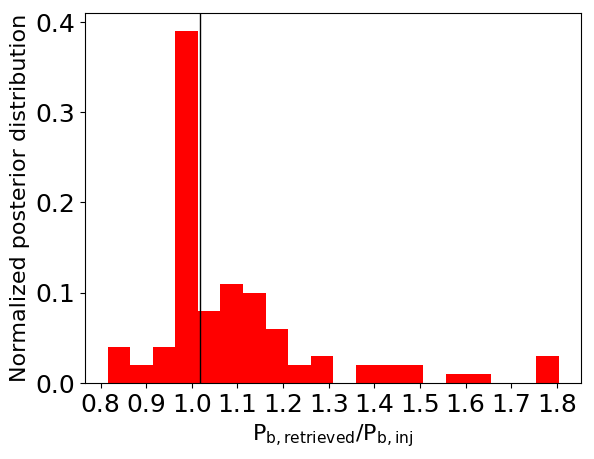}
  \caption{}
  \end{subfigure}
  \begin{subfigure}{.5\textwidth}
  \centering
  \includegraphics[width=0.85\hsize]{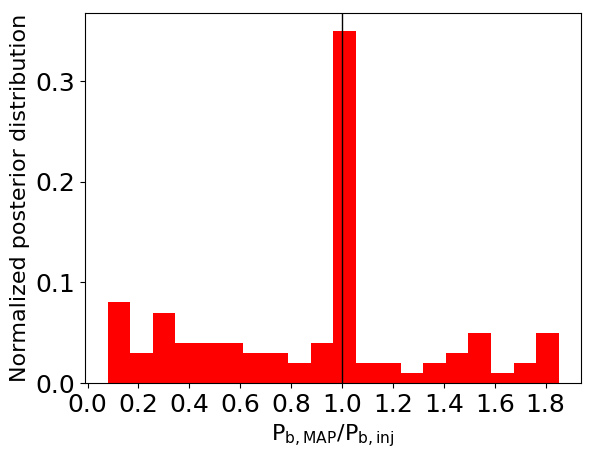}
  \caption{}
  \end{subfigure}
  \begin{subfigure}{.5\textwidth}
  \centering
  \includegraphics[width=0.85\hsize]{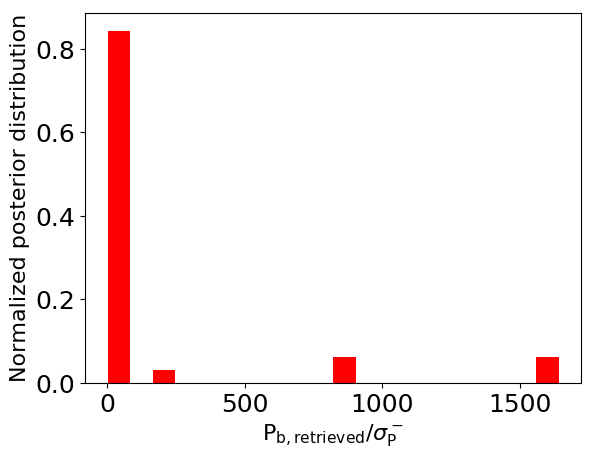}
  \caption{}
  \end{subfigure}
  \begin{subfigure}{.5\textwidth}
  \centering
  \includegraphics[width=0.85\hsize]{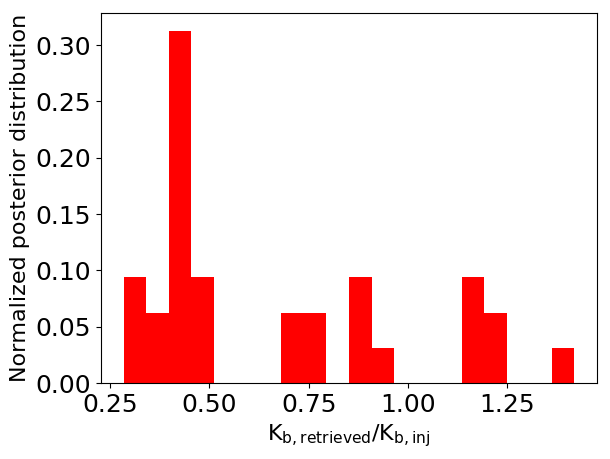}
  \caption{}
  \end{subfigure}
  \begin{subfigure}{.5\textwidth}
  \centering
  \includegraphics[width=0.85\hsize]{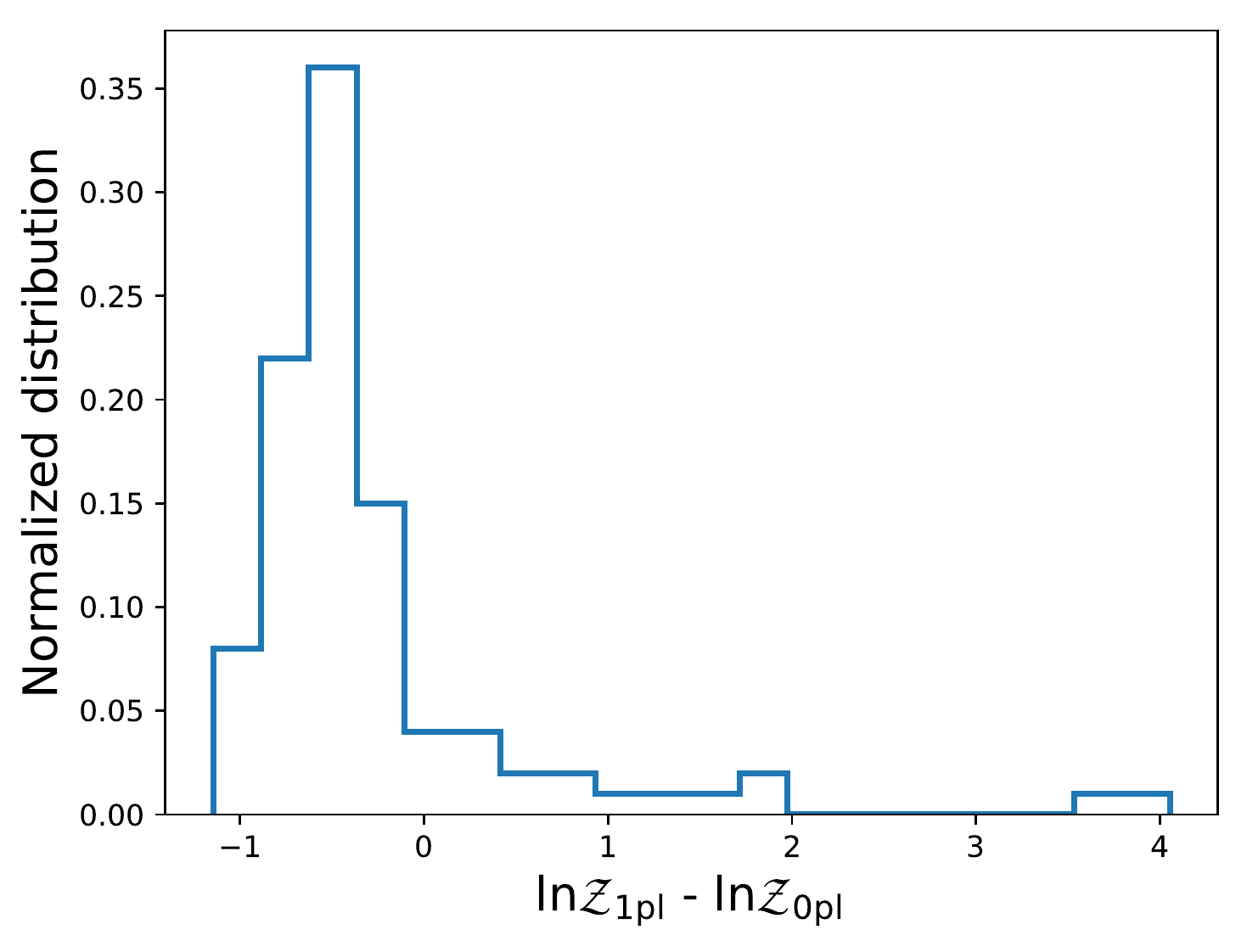}
  \caption{}
  \end{subfigure}
  \caption{\textit{Panel a}. Posterior distributions of the orbital period $P_{\rm b}$ of the injected planet for some selected mock datasets. The histogram in green corresponds to a dataset for which $P_{\rm b}$ has been recovered with high precision and accuracy. The posterior in blue corresponds to a case with a very accurate but less precise retrieved $P_{\rm b}$, while the posterior in red corresponds to a mock dataset for which the orbital period wasn't recovered.
  \textit{Panel b}. Distribution of the $P_{\rm b,\,retrieved}$/$P_{\rm b,\:inj}$ ratio between the median best-fit orbital period and the injected value, for each simulated dataset. 
  \textit{Panel c}. Same as for \textit{panel b}, but using the values of $P_{\rm b}$ corresponding to the maximum a posteriori probability (MAP) estimates.
  \textit{Panel d}. Distribution of the significance of the retrieved best-fit values $P_{\rm b}$ for the $S_{32}$ sub-sample described in the text. The significance is expressed as the ratio between the median of the posterior and the corresponding lower uncertainty, for each simulated dataset.
  \textit{Panel e}. Distribution of the ratio between the recovered and injected semi-amplitude $K_{\rm b}$ of the planetary RV signal for the 32$\%$ of the simulated datasets for which an accurate and precise estimate of $P_{\rm b}$ was recovered ($S_{32}$ sample).   
  \textit{Panel f}. Results of a model comparison analysis. The plot shows the distribution of the difference $\ln\mathcal{Z}_{1pl}$-$\ln\mathcal{Z}_{0pl}$, calculated for the 100 mock RV datasets, between the Bayesian evidences of the model including one sinusoid (i.e. one planet on a circular orbit) and the model with only a GP-term. The model including one planet is marginally more significant ($\ln\mathcal{Z}_{1pl}$-$\ln\mathcal{Z}_{0pl}>$2) only in two cases, one being that corresponding to the posterior of $P_{\rm b}$ in green shown in \textit{panel a}. 
   }
  \label{Fig:sim2results}
\end{figure*}

We analysed each mock dataset with a GP regression including a sinusoid, and with the same set-up used for the real RVs. The real $\sigma_{\rm RV}$ uncertainties were used as error bars. Figure \ref{Fig:sim2results} summarises some main outcomes of the analysis. 
Panel (a) shows examples of posterior distributions for the orbital period $P_{\rm b}$, corresponding to datasets for which $P_{\rm b}$ was recovered with high and good accuracy, and one case for which $P_{\rm b}$ was not recovered at all.  
Panel (b) shows the distribution of the $P_{\rm b,\,retrieved}$/$P_{\rm b,\:inj}$ ratio, and panel (c) shows the distribution of the $P_{\rm b,\,MAP}$/$P_{\rm b,\:inj}$ ratio. For 61$\%$ of the samples, the best-fit median $P_{\rm b,\,retrieved}$ falls within $\pm$0.5~d of the injected orbital period, that we assume as the interval corresponding to an accurate and potentially precise (and significant), detection. Nearly half of this sub-sample (corresponding to 32$\%$ of the total mock datasets, that we will call the $S_{32}$ sample for convenience), has the MAP value falling within the same interval\footnote{In other words, this is the percentage of the simulated datasets for which we recovered an accurate and reliable estimate of $P_{\rm b}$, as indicated by the MAP values.}. The percentage of the total samples for which $P_{\rm b,\,MAP}$ falls within $\pm$0.5~d of the injected orbital period is 40$\%$, thus meaning that for 8$\%$ of the samples with quite accurate MAP values we did not recover precise best-fitting values of $P_{\rm b}$. The median and mean significance of the recovered orbital period for the $S_{32}$ sample are 1.8$\sigma$ and 164$\sigma$ respectively (see panel (d) of Fig.~\ref{Fig:sim2results}).  
Panel (e) shows the distribution of the ratio between the recovered and injected semi-amplitude $K_{\rm b}$ for the sample $S_{32}$. We found that for more than half of the samples we recovered inaccurate values for $K_{\rm b}$, which are less than half the injected value.     

We also analysed each simulated dataset without including a sinusoid to model the signal due to the injected planet. We compared the Bayesian evidences $\ln\mathcal{Z}_{1pl}$ and $\ln\mathcal{Z}_{0pl}$ derived by \texttt{MultiNest} in order to assess how much the correct model (i.e. that including the planetary signal) is statistically favoured. This, in turn, gives us information on how effective our methods are at retrieving the injected signal. The result is shown in the last panel of Fig.~\ref{Fig:sim2results}. The model including one circular planetary signal is never more probable, except for two datasets only with $\ln\mathcal{Z}_{1pl}- \ln\mathcal{Z}_{0pl}>$2.

These simulations indicate that even with a large number (well over a hundred) of high-quality RVs from an instrument such as HARPS-N, reliably detecting the planet claimed by D017 to orbit \object{V830 Tau} is almost invariably going to be extremely difficult.


\section{Discussion and conclusions}
\label{sec:discussion}
After the announcement of the discovery of a HJ with the radial velocity method, the $\sim$2 Myr old star \object{V830 Tau} has become a milestone for our understanding of the formation and evolution time scales of extrasolar planets \citep{donati16,donati17}. The detection of a planet at an early stage of formation in a close-in orbit around its host ($a=0.057$ au) revealed that Jupiter-like planets can migrate inwards in less than 2 Myr. This discovery motivated an RV follow-up campaign of \object{V830 Tau} within the GAPS programme, with the main goal of improving the planetary parameters using the high-resolution HARPS-N spectrograph.

With a variability of the order of $1~\kms$ observed in the RV time series, being almost entirely due to magnetic activity, \object{V830 Tau} is among the most active young stars monitored for blind planet searches. As such, it represents \emph{a priori} a very challenging target even when using the best spectrographs currently available, and claiming the detection of even a massive HJ with high statistical significance can be very difficult. 

\begin{figure}
\hspace*{-7mm}
 \centering{
 \includegraphics[width=\hsize]{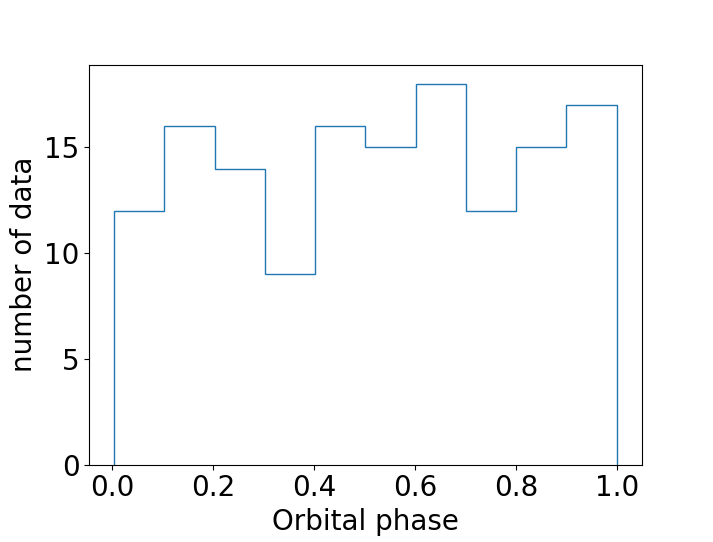}} 
   \caption{Distribution of the number of HARPS-N observations of \object{V830 Tau} according to the orbital phase of the planet announced by \citet{donati17}.}
  \label{fig:isto_phase_planet_b}%
\end{figure}

The conclusion from our analysis of HARPS-N RVs is that we cannot confirm the existence of the planetary signal attributed to V830 Tau\,b, that was claimed with a very strong statistical significance by DO17 (Bayes’ factor of $10^{8}-10^{9}$). 
To investigate the presence of the planet as carefully as possible, we analysed RVs extracted with three different pipelines, and used different methods and tools to account for the dominant activity signals and perform robust Bayesian model comparison. Our analysis also took advantage of the information embedded in simultaneous activity diagnostics from photometry and spectroscopy. 
Two of the HARPS-N RV datasets have internal errors nearly half those of DO17, but the scatter of our measurements is higher. This could be due to an increase in the level of the stellar activity since 2016, and we actually found evidence for increasing activity within the time span of our spectroscopic follow-up (Fig.~\ref{Fig:caII_halpha_actind}). This may represent a further obstacle for recovering the planetary signal, despite the quality and sampling of our data. Fig.~\ref{fig:isto_phase_planet_b} shows that our HARPS-N observations are distributed quite uniformly over the orbit of \object{V830 Tau b} -- thus our non-detection of the planet signal could not be attributed to a poor sampling.  

We also devised detailed injection-retrieval simulations based on our data (TERRA dataset) and analysis set-up to meticulously investigate our sensitivity to the presence of a planetary companion. The main results can be summarised as follows: 
\begin{itemize}
    \item After injecting the putative planetary signal into our real RV time series (Sect.~\ref{sect:simu_inj_real_data}), we could recover accurately the correct orbital period for nearly 50$\%$ of the cases ($4.9<P_{b}<5.0$), but the semi-amplitude was not retrieved with accuracy. We could recover the planet with the same high statistical significance claimed by DO17 only injecting a signal with twice the semi-amplitude of that reported in their work. However, it must be highlighted that our results depend on the adoption of quite broad, uninformative priors for the planetary model parameters in all our analyses, as expected when conducting a blind search.
    \item After injecting the planetary signal of DO17 into randomly generated RV datasets with average properties similar to the real RVs (Sect.~\ref{sect:simu_inj_varying_hyper}), we could retrieve accurate values for $P_{\rm b}$ (relying on the maximum \emph{a posteriori} probability) for $32\%$ of the realisations. For just 8 of these favourable cases, however, we could recover a precise $P_{\rm b}$ (significance of the detection $>5\sigma$), while the retrieved semi-amplitude $K_{\rm b}$ was in general not accurate for this sub-sample. For all the mock datasets, except one, model comparison based on Bayesian marginal likelihoods shows that the model including the planetary signal is never statistically favoured.
\end{itemize}
While these results do not rule out with high confidence the existence of \object{V830 Tau b}, they nonetheless clearly show that retrieving the DO17 signal (or a signal very similar to it) with high significance is far from a simple task, even given the high quality of the HARPS-N data and state-of-the-art tools and methods used for the analysis.   

\begin{figure}
\hspace*{-7mm}
 \centering{
 \includegraphics[width=\hsize]{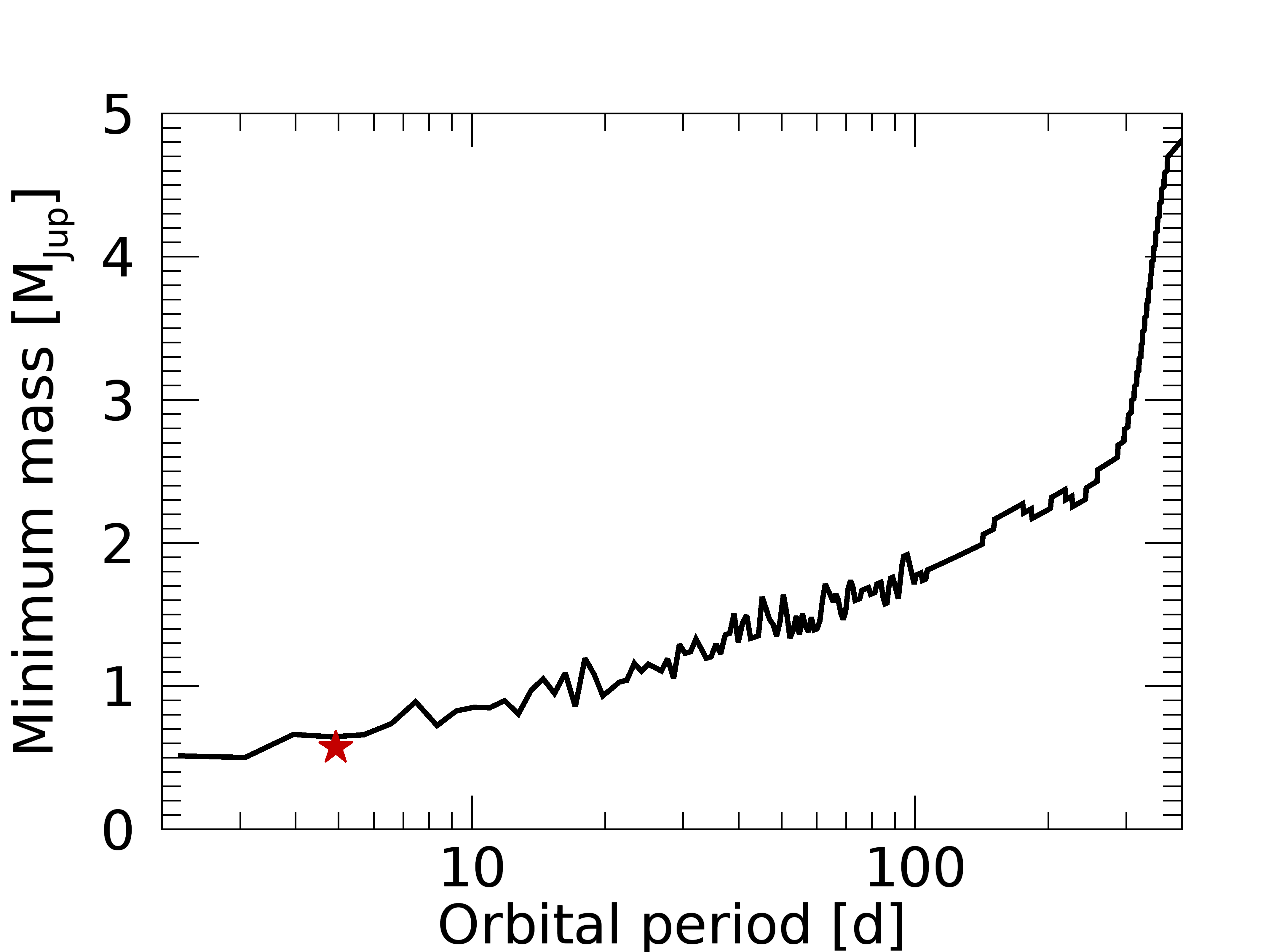}} 
   \caption{\textbf{Detection limits for planets on circular orbits based on the TERRA RV residuals for the GP model with N=0 planets. For each trial orbital period we calculated the minimum detectable value of the planet minimum mass. The red star identifies the position on the diagram of the planet announced by \citet{donati17}.}}
  \label{fig:rv_detec_lim}%
\end{figure}

We further explored the potential of our data by calculating the detection limits provided by the HARPS-N RVs. To this purpose, we derived a diagram showing the lowest minimum planetary mass we are sensitive to as a function of the planet orbital period. The calculation assumes the TERRA RV residuals of the GP model with N=0 planets, and is based on the following `frequentist' approach. First, through a bootstrap analysis we derived the power $pow_{0.1\%}$ of the GLS periodogram corresponding to a level of false alarm probability of 0.1$\%$. We then defined arrays of velocity semi-amplitudes $K$, orbital periods $P$, and orbital phases $\phi$ to generate sinusoids which simulate signals induced by a planet on a circular orbit. We adopted 10 -- 1000 \ms and 2.2 -- 440 d as the variability ranges for $K$ and $P$, respectively (with the upper limit on the orbital period equal to half the time span of our observations), and 100 linearly spaced values between 0 and 1 were generated for $\phi$. Each simulated sinusoid was injected into the original RV residuals, and we calculated the power $pow_{\rm trial}$ of the GLS periodogram at the planet orbital frequency. If, given a pair ($K$,$P$), $pow_{\rm trial}>pow_{0.1\%}$ for all the orbital phases, we consider the planet as detected. We used 1 $\msun$ for the mass of V830 Tau (from DO17) to transform the velocity semi-amplitudes to values of minimum mass for the injected planet.   
According to the results of this simplified, nonetheless illustrative calculation (Fig. \ref{fig:rv_detec_lim}), we straddle the detection limit for the $m_{\rm p}\sini_{\rm p}$=0.57 $\Mjup$ planet claimed by DO17. This is in agreement with the difficulties encountered in retrieving the planetary signal through the more complex and rigorous statistical analysis described in Sect. \ref{sect:simu_inj_real_data}.

Our work was intended as an independent investigation of the \object{V830 Tau} system using HARPS-N, and therefore we do not present here any re-analysis of the data from DO17. We cannot fully reject the reality of the 4.9-d signal claimed by DO17, but we believe the HARPS-N observations and our analyses do cast doubts on DO17's signal having a planetary origin. Further work -- new observations (even with NIR spectrographs, and possibly during epochs of lower stellar activity), more sophisticated analysis techniques and/or perhaps a better understanding of nuisance signals in existing RVs -- will clearly be needed to definitively confirm or refute the existence of \object{V830 Tau b}. This point is of crucial relevance for assessing the occurrence rate of HJs around young stars, and for understanding the formation paths and migration mechanisms that apparently might bring them to move close to its new born star in a short time scale, before the dissipation of the protoplanetary disk . 

One main conclusion of our work, is that any detection based on the RVs alone should be taken with extreme caution, and that independent re-analysis and follow-up are strongly encouraged on a case-by-case basis. Good examples of debated RV-detected HJs around young stars are represented by \object{TW Hya} and \object{Cl Tau}. The first is the closest T Tauri star to the Sun ($\sim$10 Myr old), with a candidate giant planet ($\sim$10 $M_{\rm Jup}$) detected at a separation of 0.04 au ($P\sim$3.5 d) by \cite{setiawan08}. The existence of this close-in companion was debated by \cite{huelamo08}, who concluded that the RV signal could be best explained by a long-lasting cool stellar spot on the stellar surface. 
\object{Cl Tau} is a star coeval to \object{V830 Tau} with the first HJ candidate ($P\sim$9 d) detected within the very young protoplanetary disk \citep{johnskrull16}. This detection has been recently questioned and the signal attributed instead to stellar activity \citep{donati20}. Interestingly, for \object{Cl Tau} there is evidence for ongoing giant planet formation at larger separations (10-100 au), as revealed by \cite{Clarke2018} using high-resolution imaging with the Atacama Large Millimeter/submillimeter Array (ALMA). Moreover, the existence of a HJ orbiting the older $\sim$150 Myr star \object{BD+20 1790} was ruled out by the GAPS collaboration using near-infrared RVs \citep{carleo18}, and still GAPS observations, using the combination of HARPS-N (VIS) and GIANO (NIR) RVs, enabled to exclude the existence of an HJ orbiting AD Leo (age between 25 and 300 Myr) \citep{2020A&A...638A...5C}. 
Nowadays, detecting young planets in close-in orbits with the photometric transit method remains the more secure way to ascertain their existence, nevertheless their precise characterisation with spectroscopic follow-up observations is still challenging.

\begin{acknowledgements}
We thank the anonymous referee for her/his consideration and useful suggestions.
We thank Daniele Locci (INAF-OAPa) for useful comments.
We acknowledge the support by INAF/Frontiera through the "Progetti Premiali" funding scheme of the Italian Ministry of Education, University, and Research. 
We acknowledge the computing centres of INAF - Osservatorio Astronomico di Trieste / Osservatorio Astrofisico di Catania, under the coordination of the CHIPP project, for the availability of computing resources and support. We thank Chris Sneden for providing us with the line list useful for measuring stellar parameters. VMR thanks the Royal Astronomical Society and Emmanuel College, Cambridge, for financial support.
This research has made use of the VizieR catalogue access tool, CDS, Strasbourg, France (DOI: 10.26093/cds/vizier)
\end{acknowledgements}

%
%

\bibliographystyle{aa}
\bibliography{refYO01}









\begin{appendix}

\section{Light curve measurements}

\begin{table}
\caption{STELLA \textit{V}-band photometric data. The complete table is made available in electronic form at the CDS.}             
\label{table:stellavband}      
\centering                 
\begin{tabular}{l l l}       
\hline\hline              
Time (BJD$-2\,450\,000$) & Relative flux & Relative flux error \\    
\hline                        
8535.375000 & 0.932446 & 0.003388 \\
8542.335938 & 1.038510 & 0.002648 \\
8545.402344 & 0.940533 & 0.003131 \\
8546.335938 & 0.913252 & 0.001906 \\
... & ... & ... \\
\hline                                   
\end{tabular}
\end{table}

\begin{table}
\caption{STELLA \textit{I}-band photometric data. The complete table is made available in electronic form at the CDS.}             
\label{table:stellaiband}      
\centering                 
\begin{tabular}{l l l}       
\hline\hline              
Time (BJD$-2\,450\,000$) & Relative flux & Relative flux error \\    
\hline                        
8539.527344 & 1.036580 & 0.004056 \\
8551.406250 & 0.924861 & 0.003557 \\
8565.347656 & 0.918548 & 0.003289 \\
8566.347656 & 1.080230 & 0.003108 \\
... & ... & ... \\
\hline                                   
\end{tabular}
\end{table}

\section{Spectroscopic activity indexes}

\begin{table}
\caption{Activity diagnostics extracted from the HARPS-N spectra. For the FWHM and BIS we assume uncertainties equal to twice the internal errors $\sigma_{\rm RV_{\rm DRS}}$ \citep{santerne15}. The complete table is made available in electronic form at the CDS.}          
\label{table:actinddata}      
\centering                 
\begin{tabular}{ccccccc}       
\hline\hline              
Time & FWHM & BIS & H-$\alpha$ & $\sigma_{\rm H-\alpha}$ & CaII H$\&$K & $\sigma_{\rm CaII H\&K}$ \\    
(BJD$-2\,450\,000$) & ($\ms$) & ($\ms$) &  & & & \\ 
\hline                        
8044.662344 & 46562.3  & 236.5   &  0.981 & 0.002 & 3.783 & 0.057 \\        
8050.690209 & 42414.4. & 1290.3  &  0.795 & 0.002 & 3.236 & 0.084 \\        
8051.628520 & 43315.3  & 241.0   &  0.918 & 0.002 & 3.726 & 0.076 \\       
8052.685539 & 42600.9  & -2403.1 &  0.992 & 0.002 & 3.661 & 0.072 \\        
...         & ...      & ...      & ...   & ...   & ...  & ...  \\
\hline                                   
\end{tabular}
\end{table}

\section{Radial velocity measurements}

\begin{table}
\caption{Radial velocities extracted with TERRA and DRS pipelines, and the template-free technique of \cite{rajpaul20}, as described in Sect.~\ref{sec:rvextraction}. The average of the systemic velocity was subtracted from the DRS data. The complete table is made available in electronic form at the CDS.}          
\label{table:rvdata}      
\centering                 
\begin{tabular}{ccccccc}       
\hline\hline              
Time & RV$_{\rm TERRA}$ & $\sigma_{\rm RV_{\rm TERRA}}$ & RV$_{\rm DRS}$ & $\sigma_{\rm RV_{\rm DRS}}$ & RV$_{\rm Rajpaul+20}$ & $\sigma_{\rm RV_{\rm Rajpaul+20}}$ \\    
(BJD$-2\,450\,000$) & ($\ms$) & ($\ms$) & ($\ms$) & ($\ms$) & ($\ms$) & ($\ms$) \\ 
\hline                        
8044.662344 & 136.1   & 19.0 & 260.1   & 21.4 &  253.1 & 43.9\\        
8050.690209 & -1449.7 & 23.5 & -1682.6 & 22.9 & -1550.8 & 47.8\\        
8051.628520 & 67.9    & 21.5 & 328.9   & 22.9 & 193.5 & 53.1\\       
8052.685539 & 627.9   & 29.3 & 1479.2  & 24.2 & 1061.8 & 45.0\\        
...         & ...     & ...  & ...     & ...  & ... & ...\\
\hline                                   
\end{tabular}
\end{table}

\section{Radial velocity analysis}

\begin{figure*}
   \centering
   \includegraphics[width=\hsize]{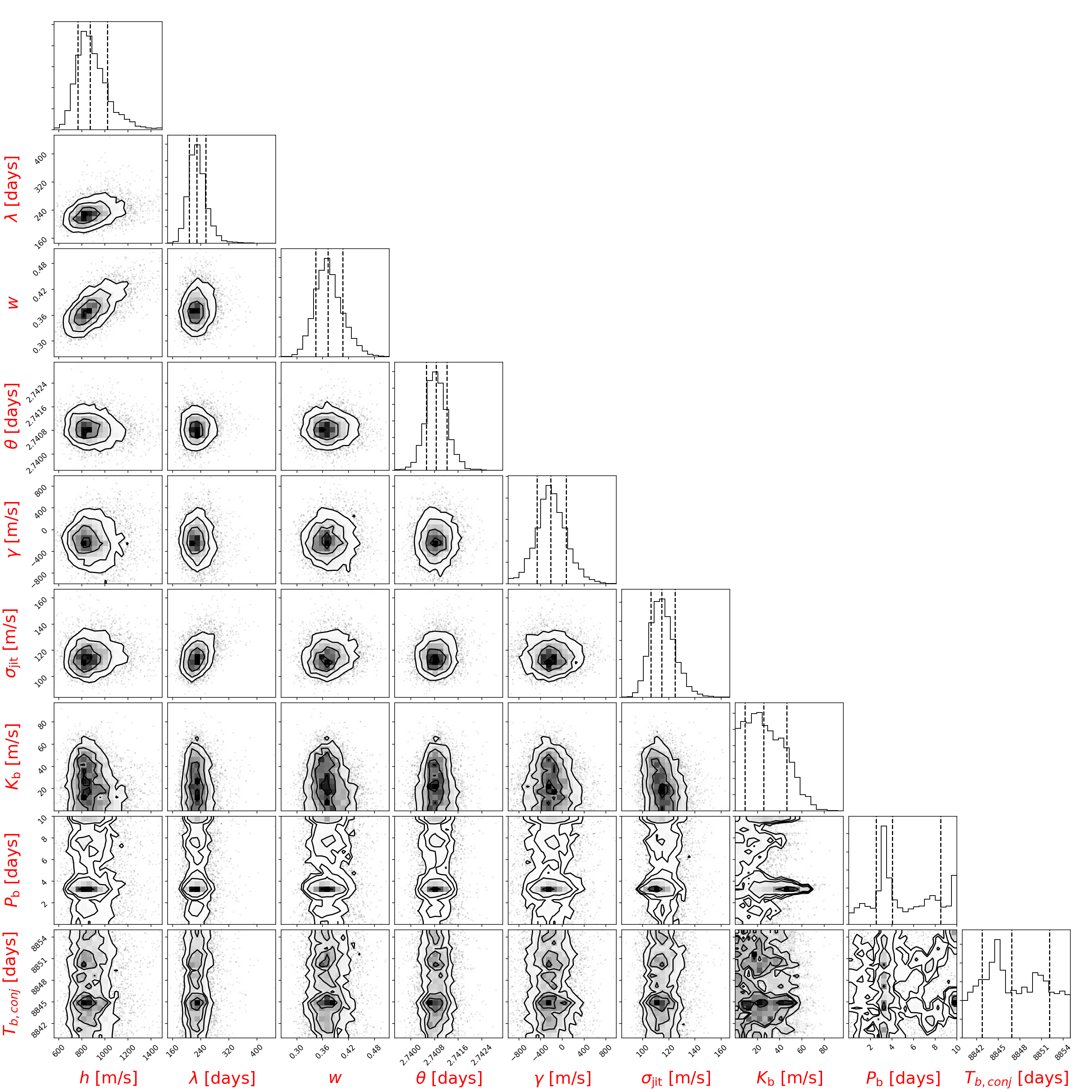}\\
      \caption{Posterior distributions of the free (hyper-)parameters of the GP+1 circular planet regression (Table~\ref{tab:percentilesgp}). The corresponding RV dataset is that extracted with TERRA pipeline.}
         \label{Fig:corneplotterragp1p}
\end{figure*}

\end{appendix}

\end{document}